\documentclass[a4paper, 11pt]{article}   	

\usepackage{geometry}                		
\usepackage{epsfig}
\usepackage{graphicx}				
\usepackage{amssymb}
\usepackage{amsmath}
\usepackage{bm}
\usepackage{mathtools}
\usepackage{flexisym}
\usepackage{color}

\usepackage{comment}				

\usepackage[T1]{fontenc} 

\begin{document}

\begin{center}
\LARGE\bf Phenomenology of SUGRA Extensions of the Starobisnky Model
\end{center}

\footnotetext{\hspace*{-0cm}\footnotesize MAN Ping Kwan, Ellgan, E-mail:  ellgan101@akane.waseda.jp}

\begin{center}
\rm MAN Ping Kwan, Ellgan$^{\rm a)\dagger}$
\end{center}

\begin{center}
\begin{footnotesize} \sl
${Department \; of \; Pure \; and \; Applied \; Physics, \; Waseda \; University}^{\rm a)}$ \\
\end{footnotesize}
\end{center}

\section{Abstract}
\noindent We analyze BI-extended model in a complete form and compare the predictions with that of Starobinsky model. Under the parameter constraints in Planck 2018, we find that the dynamics of the whole inflation process described by BI-extended and Starobinsky models are nearly the same, even though there are some differences in the regions out of inflation. We also find the scales of parameters in BI-extended model and inflaton values at the first horizon crossing required to implement inflation. The changes of $\left( {n}_{s}, r \right)$ fingerprints of BI-extended model and that of evolutions of inflaton field due to the variations of relevant parameters are also investigated. 

\section{Introduction}
\begin{table}[h]
\begin{center}
\begin{tabular}{ |c|c|c|c| }
\hline
$\text{Slow-roll parameters}$ & $\text{Range(s)}$ & $\text{Spectral indices}$ & $\text{Range(s)}$ \\
\hline
${\epsilon}_{V}$ & $<0.004$ & ${n}_{s} - 1$ & $[-0.0423, -0.0327]$ \\
\hline
${\eta}_{V}$ & $[-0.021, -0.008]$ & ${\alpha}_{s} := \frac{d {n}_{s}}{d \ln{k}}$ & $[-0.008, 0.012]$ \\
\hline
${\xi}_{V}$ & $[-0.0045, 0.0096]$ & ${\beta}_{s} := \frac{d^{2} {n}_{s}}{d \ln{k}^{2}}$ & $[-0.003, 0.023]$ \\
\hline
${H}_{\text{hc}}$ & $< 2.7 \times {10}^{-5} {M}_{\text{pl}}$ & ${V}_{\text{hc}}$ & $< \left( 1.7 \times {10}^{16} \; \text{GeV} \right)^{4}$ \\
\hline
\end{tabular}
\end{center}
\caption{Slow roll potential parameters and spectral indices in Planck 2018}
\label{table:Planck data 2018 slow roll potential parameters and spectral indices}
\end{table}

\noindent Cosmological inflation is a powerful solution to the flatness and horizon problems at the beginning of the standard Big Bang scenario \cite{{Inflationary universe: A possible solution to the horizon and flatness problem}, {A new type of isotropic cosmological models without singularity}, {First-order phase transition of a vacuum and the expansion of the Universe}}. The constraints of observation data of Planck 2018 are listed in Table \ref{table:Planck data 2018 slow roll potential parameters and spectral indices}\footnote{Note that ${V}_{\text{hc}}$ in Table \ref{table:Planck data 2018 slow roll potential parameters and spectral indices} is the scalar power spectrum at the first horizon crossing given by
\begin{equation}
{V}_{\text{hc}} = \frac{3 {\pi}^{2}}{2} r {M}^{4}_{\text{pl}} \left. {A}_{s} \right|_{\text{hc}}, \\
\end{equation}
\noindent where $\left. {A}_{s} \right|_{\text{hc}}$ is the scalar power spectrum evaluated at the first horizon crossing. 
}. All inflation models naturally provide the primordial density fluctuation and curvature perturbation, which can be observed by cosmic microwave (CMB) background \cite{{Planck 2013 results. XXII. Constraints on inflation}, {Planck 2018 results. X. Constraints on inflation}}. So far many models listed in \cite{{Planck 2018 results. X. Constraints on inflation}, {Encyclopaedia Inflationaris}} can satisfy the observation. In particular, Starobinsky model (also called $R^2$ inflation model), motivated by modified gravity \cite{{Quantum Fluctuation and Nonsingular Universe}, {The Perturbation Spectrum Evolving from a Nonsingular Initially de Sitter Cosmology and the Microwave Background Anisotropy}}, provides the most promising prediction. This triggers many discussions about properties of Starobinsky models during and after inflation like \cite{1705.11098} and the counterparts of Starobinsky like models such as \cite{Nearly Starobinsky inflation from modified gravity} and \cite{Spotting deviations from R2 inflation}, and the dynamics of the extensions of Starobinsky model such as \cite{1504.01772} and \cite{1708.08346}. \\

\noindent On the other hand, it is vital to integrate those inflation models to high energy theories like supergravity (SUGRA). So far, SUGRA, which is a locally supersymmetric theory, has been an appropriate model to unify gravity with particle physics beyond the Standard Model of elementary particles and beyond the Standard Model of cosmology \cite{1807.08394}. In particular, the $F$-term potential in SUGRA provides many insights in $\alpha$ attractor property, which is describing the connection between Starobinsky model and various power-law model in the graph of tensor-to-scalar ratio against spectral index \cite{{1311.0472},{1607.08854}, {1611.00738}, {1612.01126}, {1707.05830}, {1712.09693}}. Although there is an obstacle called $\eta$ problem resulting from the factor ${e}^{\frac{K}{{M}^{2}_{\text{pl}}}}$ in the $F$-term potential where $K$ denotes K\"ahler potential \cite{False Vacuum Inflation with Einstein Gravity}, it was solved by making a proper choice in super-potential and K\"ahler potential as shown in \cite{Natural chaotic inflation in supergravity} and \cite{General inflaton potentials in supergravity}. Another solution is discussed in \cite{Minimal supergravity models of inflation} and \cite{Naturalness and chaotic inflation in supergravity models of inflation}, which considers the scalar component of a massive $U \left( 1 \right)$ vector multiplet as an inflaton field. This can avoid the occurrence of $\eta$ problem and realize inflation in a simple way. \\

\noindent It is natural to consider the UV completion of SUGRA inflation models. One possibility is to adopt Dirac-Born-Infeld (DBI) action, which carries zero-mode vector fields attached to the $D$-branes \cite{{Foundations of the new field theory}, {An extensible model of the electron}}. The supersymmetric (SUSY) version of DBI action was studied in \cite{Supersymmetric Born-infeld Lagrangians} and \cite{hep-th/9608177}. The matter coupled DBI type action including a matter charged under $U \left( 1 \right)$ symmetry is discussed in \cite{1504.01221}. The action of the massive vector multiplet is extended to the DBI type action, which can be constructed by the DBI action of a massless vector multiplet coupled to a Stueckelberg multiplet with $U \left( 1 \right)$ symmetry \cite{1505.02235}, leading to the DBI extension of Starobinsky model and the corresponding dual action\footnote{The duality described here is originated from \cite{hep-th/9608177}. }, with the inflaton field being stored in the lowest scalar component of a massive vector multiplet. This model was also studied in \cite{1807.08394}, and the effective scales are evaluated under the approximation of the square root term and compared with that of Starobinsky model, with the assumption that the inflaton field is induced from the scalar curvature by the Legendre transformation instead of the scalar field of the massive vector multiplet. \\

\noindent Thus, in this paper, we study the version of \cite{1807.08394} in a complete form\footnote{We assume the inflaton field is from the scalar curvature instead of the scalar field of the massive vector multiplet. } to see how all the higher order terms contribute to the predictions. For the rest of this paper, we call the model we are going to study as "BI-extended model", although there are many meanings for "BI" like "Brane Inflation"\cite{Brane inflation} and DBI inflation \cite{Scalar speed limits and cosmology: Acceleration from D-cceleration}. In section \ref{Basic Formalism for studying inflation with $F(R)$ model}, we introduce the basic formalism of obtaining the potential from a general $F \left( R \right)$ term in a Lagrangian density. After a short review in Starobinsky model in section \ref{A Short Review: Starobinsky Model}, we analyze the BI-extended model in section \ref{The BI-extended Model}. In section \ref{Numerical Calculation}, we show the possible parameters and inflaton values at the first horizon crossing ${\phi}_{\text{hc}}$ in Starobinsky model and BI-extended model based on the observation constraints listed in Table \ref{table:Planck data 2018 slow roll potential parameters and spectral indices} and how BI-extended model changes as $g$ and ${\beta}$ vary. In section \ref{Evolutions of inflaton field}, we demonstrate the evolutions of inflaton field as cosmic time $t$ runs in different choices of $g$ and ${\beta}$. Finally, we discuss our results in section \ref{Discussions} and summarize them in section \ref{Conclusions}. 

\section{Basic Formalism for studying inflation with $F(R)$ model}
\label{Basic Formalism for studying inflation with $F(R)$ model}
\noindent In this section, we are going to review the basic derivation of $F(R)$\footnote{Note that the mass scale of $F \left( R \right)$ is $\left[ F \left( R \right) \right] = {M}^{2}_{\text{pl}}$, where ${M}_{\text{pl}}$ is the reduced Planck mass. } theory\footnote{For general review, please refer to \cite{1011.0544}. }. Note that the Lagrangian is given by

\begin{equation}
{S}_{\text{Jordan}} = \frac{{M}^{2}_{\text{pl}}}{2} \int d^{4} x \sqrt{-\tilde{g}} F(\tilde{R})
\end{equation}
\noindent where $\tilde{{g}}_{{\mu}{\nu}}$ is the metric in the Jordan frame\footnote{In this paper, notations with a tilde mean the corresponding quantities evaluated in the Jordan frame. }. To get the equation of motion (E.O.M.), we vary the action with respect to the metric $\tilde{g}^{{\mu}{\nu}}$ to produce

\begin{equation}
F'(\tilde{R}){\tilde{R}}_{{\mu}{\nu}} - \frac{1}{2} F(\tilde{R}) \tilde{g}_{{\mu}{\nu}} - \left[ {\triangledown}_{\mu} {\triangledown}_{\nu} - \tilde{g}_{{\mu}{\nu}} {\square} \right] F'(\tilde{R}) = \frac{1}{{M}^{2}_{\text{pl}}} \tilde{T}_{{\mu}{\nu}}, \\
\end{equation}
where $F''(\tilde{R}) \neq 0$ and $\tilde{T}_{{\mu}{\nu}}$ is the energy-momentum tensor in the Jordan frame. Also, by taking Legendre Transformation and introducing the new scalar field $\chi$, we have
\begin{equation}
F(\tilde{R})=F'(\chi)(\tilde{R}-\chi)+F(\chi). \\
\end{equation}

\noindent In order to change it from Jordan frame to Einstein frame, we adopt the following conformal transformation

\begin{equation}
\tilde{g}_{{\mu}{\nu}} = {\Omega}^{2}(x) {g}_{{\mu}{\nu}} \\
\end{equation}
where ${\Omega} \left( x \right)$ is a function of the space-time coordinates. Note that the Christoffel symbols, Riemann tensor, Ricci tensor and Ricci scalar in the Jordan frame can be written in terms of the counterparts in the Einstein frame as follows
\begin{equation}
\begin{split}
\tilde{{\Gamma}}^{\rho}_{{\mu}{\nu}} =& {\Gamma}^{\rho}_{{\mu}{\nu}} + \left( {\delta}^{\rho}_{\mu} {\partial}_{\nu}\ln{\Omega} + {\delta}^{\rho}_{\nu} {\partial}_{\mu}\ln{\Omega} - {g}_{{\mu}{\nu}} {\partial}^{\rho} \ln{\Omega} \right), \\
{\tilde{R}}^{a}_{\;\;bcd} =& {R}^{a}_{\;\;bcd} - 2\left( {\delta}^{a}_{{c} {\delta}^{e}_{d}} {\delta}^{f}_{b} - {g}_{b[c} {\delta}^{e}_{d]} {g}^{af} \right) \frac{{\triangledown}_{e} {\triangledown}_{f} {\Omega}}{\Omega}, \\
& + 2 \left( 2 {\delta}^{a}_{[c} {\delta}^{e}_{d]} {\delta}^{f}_{b} - 2 {g}_{b[c} {\delta}^{e}_{d]} {g}^{af} + {g}_{b[c} {\delta}^{a}_{d]} {g}^{ef} \right) \frac{({\triangledown}_{e}{\Omega})({\triangledown}_{f}{\Omega})}{{\Omega}^{2}}, \\
{\tilde{R}}_{bc} =& {R}_{bc} - \left[ (d-2) {\delta}^{f}_{b} + {g}_{bc} {g}^{ef} \right] \frac{{\triangledown}_{e} {\triangledown}_{f} {\Omega}}{{\Omega}} + \left[ 2 (d-2) {\delta}^{f}_{b} {\delta}^{e}_{c} - (d-3) {g}_{bc} {g}^{ef} \right] \frac{({\triangledown}_{e} {\Omega})({\triangledown}_{f} {\Omega})}{{\Omega}^{2}}, \\
\tilde{R} =& \frac{R}{{\Omega}^{2}} - 2(d-1) {g}^{ef} \frac{{\triangledown}_{e} {\triangledown}_{f} {\Omega}}{{\Omega}^{3}} - (d-1)(d-4){g}^{ef} \frac{({\triangledown}_{e}{\Omega}) ({\triangledown}_{f}{\Omega})}{{\Omega}^{4}}. \\
\end{split}
\end{equation}
\noindent where $d$ is the dimension of the space-time. Hence, in four dimension ($d=4$), we can change all the variables from the Jordan frame to the Einstein frame
\begin{equation}
\begin{split}
{S}_{\text{Einstein}} &= \frac{{M}^{2}_{\text{pl}}}{2} \int d^{4} x \sqrt{-g} \left\{ R - \frac{3}{2} \left( \frac{F''({\chi})}{F'({\chi})} \right)^{2} {g}^{{\mu}{\nu}} \partial_{\mu} {\chi} \partial_{\nu} {\chi} - \left[ \frac{{\chi}}{F'({\chi})} - \frac{F({\chi})}{F'({\chi})^{2}} \right] \right\} \\
&= \int d^{4} x \sqrt{-g} \left[ \frac{{M}^{2}_{\text{pl}}}{2} R - \frac{1}{2} {g}^{{\mu}{\nu}} \partial_{\mu} {\phi} \partial_{\nu} {\phi} - V(\chi) \right]
\end{split}
\end{equation}

\noindent where 

\begin{equation}
\label{inflaton field redefinition}
{\Omega}^{-2}(x) = F' \left(\chi \right) ={e}^{\sqrt{\frac{2}{3}} \frac{\phi}{{M}_{\text{pl}}}}, \\
\end{equation}

\noindent and the potential 
\begin{equation}
V(\chi) = \frac{{M}^{2}_{\text{pl}}}{2} \left[ \frac{\chi}{F'({\chi})} - \frac{F({\chi})}{F'({\chi})^{2}} \right]. \\
\end{equation}
\noindent The first derivative of the potential with respect to $\chi$ is
\begin{equation}
\frac{d V \left( {\chi} \right)}{d {\chi}} = \left( \frac{{M}^{2}_{\text{pl}}}{2} \right) \frac{F'' \left( {\chi} \right) \left[ 2 F \left( {\chi} \right) - {\chi} F' \left( {\chi} \right) \right]}{F' \left( {\chi} \right)^{3}}. 
\end{equation}
\noindent Since $\frac{d V}{d {\phi}} = \frac{d V}{d {\chi}} \frac{d {\chi}}{d {\phi}}$ and 
\begin{equation}
\frac{d {\chi}}{d {\phi}} = \sqrt{\frac{2}{3}}\frac{1}{{M}_{\text{pl}}} \frac{F' \left( {\chi} \right)}{F'' \left( {\chi} \right)}, \\
\end{equation} 
\noindent we have
\begin{equation}
\frac{d V}{d {\phi}} = \frac{{M}_{\text{pl}}}{\sqrt{6}} \frac{2 F \left( {\chi} \right) - {\chi} F' \left( {\chi} \right)}{F' \left( {\chi} \right)^{2}}, \\
\end{equation}
\noindent and $R$ and $F$ in terms of the potential $V$
\begin{equation}
\begin{split}
R =&\; \left( \frac{\sqrt{6}}{{M}_{\text{pl}}} \frac{d V}{d {\phi}} + \frac{4 V}{{M}^{2}_{\text{pl}}} \right) {e}^{\sqrt{\frac{2}{3}} \frac{\phi}{{M}_{\text{pl}}}}, \\
F =&\; \left( \frac{\sqrt{6}}{{M}_{\text{pl}}} \frac{d V}{d {\phi}} + \frac{2 V}{{M}^{2}_{\text{pl}}} \right) {e}^{ 2 \sqrt{\frac{2}{3}} \frac{\phi}{{M}_{\text{pl}}}}. \\
\end{split}
\end{equation}
\noindent We adopt the Friedmann-Robertson-Walker (FRW) universe with the metric
\begin{equation}
ds^{2} = -dt^{2} + a(t)^{2} d\boldsymbol{x}^{2} = -dt^{2} + a(t)^{2} \left( d{x}^{2} + d{y}^{2} + + d{z}^{2} \right), \\ 
\end{equation}
\noindent where $a=a(t)$ is the scale factor and $t$ is the cosmic time. Since the Einstein equation in the Einstein frame is given by
\begin{equation}
{R}_{{\mu}{\nu}} - \frac{1}{2} {g}_{{\mu}{\nu}} R = \frac{1}{{M}^{2}_{\text{pl}}} {T}_{{\mu}{\nu}}, \\
\end{equation}
\noindent where ${T}_{{\mu}{\nu}}$ is the energy-momentum tensor in the Einstein frame given by
\begin{equation}
{T}_{{\mu}{\nu}} = {\partial}_{\mu} {\phi} {\partial}_{\nu} {\phi} - {g}_{{\mu}{\nu}} \left[\ \frac{1}{2} {\partial}^{\sigma} {\phi} {\partial}_{\sigma} {\phi} + V \left( {\phi} \right) \right], \\
\end{equation}
\noindent if we take homogeneous scalar field ${\phi} \left( t, \boldsymbol{x} \right) = {\phi} \left( t \right)$, the equation of motion of the scalar field (inflaton field) in the Einstein frame is
\begin{equation}
\ddot{\phi}(t) + 3 H \dot{\phi}(t) + V'(\phi) = 0, \\
\end{equation}
\noindent and the $00$ and $ij$ components of the Einstein equation in the Einstein frame are given by
\begin{equation}
\label{Einstein equations}
\begin{split}
3 {M}_{\text{pl}}^{2}{H}^{2} &= \frac{1}{2} \dot{\phi}^{2} + V(\phi), \\
\dot{H} &= - \frac{1}{2 {M}^{2}_{\text{pl}}} \dot{\phi}^{2}, \\ 
\end{split}
\end{equation}
\noindent where $H \left( t \right) = \frac{\dot{a} \left( t \right)}{{a} \left( t \right)}$ is the Hubble parameter. 

\section{A Short Review: Starobinsky Model}
\label{A Short Review: Starobinsky Model}
\noindent $F \left( R \right)$ of the Starobinsky model is given by
\begin{equation}
\label{F term Starobinsky}
{F}_{\text{Starobinsky}} \left( R \right) = R + 4 g^2 R^2, \\
\end{equation}
\noindent where $g \in \mathbb{R}^{+}$. After some algebra, we can obtain the potential in terms of inflaton field $\phi$
\begin{equation}
\label{Starobinsky potential}
{V}_{\text{Starobinsky}} \left( {\phi} \right) = \frac{{M}^{2}_{\text{pl}}}{32 g^2} \left( 1 - {e}^{- \sqrt{\frac{2}{3}} \frac{\phi}{{M}_{\text{pl}} } } \right)^{2}, \\
\end{equation}

\noindent Originally, Starobinsky model was obtained by taking the lowest order of scalar curvature correction as in \cite{The Perturbation Spectrum Evolving from a Nonsingular Initially de Sitter Cosmology and the Microwave Background Anisotropy}. But in the recent years, there have been many motivations to obtain this model, one of which is $\mathcal{N}=1$ SUGRA \cite{{1505.02235}, {1307.1137}}. In the following two subsections, we can see how the conformal SUGRA action can form the Starobinsky model, where the inflaton field comes from the lowest scalar component of a massive vector multiplet, and the dual of this action, where the bosonic part contains Starobinsky $F\left( R \right)$ term. 

\subsection{From the conformal SUGRA, \dots}
\noindent The conformal SUGRA action is given by \cite{{1307.1137}, {1505.02235}}
\begin{equation}
\label{conformal SUGRA}
S = \left[ \frac{1}{2} {S}_{0} \bar{{S}_{0}} \Phi \left( {\Lambda} + \bar{\Lambda} + g' V \right) \right] - \frac{1}{4} \left[ \mathcal{W}^{\alpha} \left( V \right) \mathcal{W}_{\alpha} \left( V \right) \right]_{F}, \\
\end{equation}
\noindent where ${S}_{0}$ is the chiral compensator, ${\Lambda}$ is a Stueckelberg chiral multiplet, $V$ is a vector multiplet, $g'$ is the gauge coupling, ${\Phi} \left( x \right)$ is an arbitrary real function of $x$, ${\mathcal{W}}^{\alpha}$ is the field strength super-multiplet and $\left[ \dots \right]_{D,F}$ are the super-conformal $D$ and $F$ term density formulae respectively \cite{Linear versus non-linear supersymmetry in general}. After taking super-conformal gauge standard and integrating out all the auxiliary fields, we obtain the boson part of the Lagrangian in the Einstein frame as

\begin{equation}
\mathcal{L}_{B} = \sqrt{-g} \left[ \frac{  {M}^{2}_{\text{pl}} }{2} R - \frac{1}{4} \mathcal{F}_{{\mu}{\nu}} \mathcal{F}^{{\mu}{\nu}} - \frac{ {g'} ^2}{2} J' \left( C \right)^{2} - \frac{1}{2} J'' \left( C \right) {\partial}_{\mu} C {\partial}^{\mu} C - \frac{ {g'}^2 }{2} J'' \left( C \right) \left( {A}_{\mu} - \frac{1}{g} {\partial}_{\mu} {\theta} \right)^{2} \right], \\
\end{equation}
\noindent where $C = \text{Re}{\Lambda}$, ${\theta} = \text{Im} {\Lambda}$ and ${\Lambda}$ is a Stueckelberg chiral and the prime on $J$ is the derivatives with respect to $C$, ${A}_{\mu}$ is the vector component of $V$, ${F}_{{\mu}{\nu}}$ is the field strength of ${A}_{\mu}$ and $R$ denotes the Ricci scalar. Taking the $U \left( 1 \right)$ gauge as ${\theta} = 0$, ${A}_{\mu} = 0$ and $J = - \frac{3}{2} \ln{\left( - \frac{1}{3} C {e}^{C} \right)}$ and $\frac{\phi}{{M}_{\text{pl}}} =  \sqrt{\frac{3}{2}} \ln{\left( - C \right)}$, the Lagrangian becomes

\begin{equation}
\mathcal{L}_{B} = \sqrt{-g} \left[ \frac{{M}^{2}_{\text{pl}}}{2} R- \frac{1}{2} {\partial}_{\mu} {\phi} {\partial}^{\mu} {\phi} - \frac{9 {g'}^2}{8} \left( 1 - {e}^{- \sqrt{\frac{2}{3}} \frac{\phi}{ {M}_{\text{pl}} } } \right)^{2} \right], \\
\end{equation}

\noindent which gives the Starobinsky model by taking $3 g' = \frac{ {M}_{\text{pl}} }{2 g}$.

\subsection{From the new minimal SUGRA, \dots}
\noindent The Starobinsky action in the new minimal SUGRA is given by \cite{{1307.1137}, {1505.02235}}
\begin{equation}
\label{Starobinsky action in the new minimal SUGRA}
{S}_{\text{dual}} = \left[ \frac{3}{2} {L}_{0} {V}_{R} \right]_{D} - {\gamma}' \left[ \mathcal{W}^{\alpha} \left( {V}_{R} \right) \mathcal{W}_{\alpha} \left( {V}_{R} \right) \right]_{F}. \\
\end{equation}

\noindent where
\begin{equation}
{V}_{R} = \ln{\left( \frac{{L}_{0}}{\left| \mathcal{S} \right|^{2}} \right)} = {\Lambda} + \bar{\Lambda} + V, \\
\end{equation}
\noindent is the dual transformation. $\mathcal{S}$ is a chiral multiplet, ${V}_{R}$ is a real multiplet, ${L}_{0}$ is a real linear compensator and ${\gamma}'$ is a real constant. After taking this dual transformation, one can obtain the dual action

\begin{equation}
{S}_{\text{dual}} = \left[ \frac{3}{2} {L}_{0} {V}_{R} \right]_{D} - {\gamma}' \left[ \mathcal{W}^{\alpha} \left( {V} \right) \mathcal{W}_{\alpha} \left( {V} \right) \right]_{F}. \\
\end{equation}

\noindent After the field redefinition ${\Lambda} \rightarrow {\Lambda}/2$, $V \rightarrow g' V / 2$, $\mathcal{S} \rightarrow {S}_{0}/ \sqrt{3}$, we obtain
\begin{equation}
{S}_{\text{dual}} = \left[ \frac{1}{2} \left| {S}_{0} \right|^{2} \left( \frac{1}{2} \left( {\Lambda} + \bar{\Lambda} + g' V \right) \right) \text{exp} \left( \frac{1}{2} \left( {\Lambda} + \bar{\Lambda} + g' V \right) \right) \right] - \frac{ {g'}^2 {\gamma}' }{4} \left[ \mathcal{W}^{\alpha} \left( {V} \right) \mathcal{W}_{\alpha} \left( {V} \right) \right]_{F}. \\
\end{equation}
\noindent By taking ${\gamma}' = {g'}^{-2}$ and ${\Phi} \left( x \right) = x {e}^{x}$, we obtain Eq.(\ref{conformal SUGRA}). After the super-conformal gauge fixings, the bosonic part of the Lagrangian $\mathcal{L}$ of Eq.(\ref{Starobinsky action in the new minimal SUGRA}) is 
\begin{equation}
\frac{1}{\sqrt{-g}} \left. \mathcal{L} \right|_{B} = \frac{{M}_{\text{pl}}^{2} }{2} R + \frac{{M}_{\text{pl}}^{4} {\gamma}'}{18} R^2 - {\gamma}' \mathcal{F}^{\left( R \right)}_{{\mu}{\nu}} \mathcal{F}^{\left( R \right) {\mu}{\nu}} - \frac{3}{2} {A}^{\left( R \right)}_{\mu} {B}^{\mu} + \left( \frac{3}{4} + \frac{2 {\gamma}' }{3} R \right) {B}_{\mu} {B}^{\mu} + {\gamma}' \left( {B}_{\mu} {B}^{\mu} \right)^{2}, \\
\end{equation}
\noindent where ${A}^{\left( R \right)}_{\mu}$ is the vector component of ${V}_{R}$, $\mathcal{F}^{\left( R \right)}_{{\mu}{\nu}} = {\partial}_{\mu} {A}^{\left( R \right)}_{\nu} - {\partial}_{\nu} {A}^{\left( R \right)}_{\mu}$ and ${B}_{\mu}$ is an auxiliary vector component of ${L}_{0}$. By taking ${A}^{\left( R \right)}_{\mu} = {B}_{\mu} = 0$, we obtain the Starobinsky $F\left( R \right)$ term $F \left( R \right) = {R} + {M}^{2}_{\text{pl}} {\gamma}' R^2/9$ with an aid of ${\gamma}' = {g'}^{-2}$ and $3 g' = \frac{{M}_{\text{pl}}}{2g}$. 

\section{The BI-extended model}
\label{The BI-extended Model}
\noindent Now we are going to consider BI-extended model
\begin{equation}
\label{BI-extended model}
{F}_{\text{BI}}(R) = R + \frac{2g^2}{3 {\beta}} \left( \sqrt{1+12 {\beta} R^2} - 1 \right), \\
\end{equation}
\noindent where\footnote{Based on the fact that $\left[ F \left( R \right) \right] = {M}^{2}_{\text{pl}}$, we can know that the mass scales of $R$, $\beta$ and $g$ are $\left[ R \right] = {M}^{2}_{\text{pl}}$, $\left[ \beta \right] = {M}^{-4}_{\text{pl}}$ and $\left[ g \right] = {M}^{-1}_{\text{pl}}$. }
\begin{equation}
\label{e and MBI}
g = \frac{1}{e {M}_{\text{pl}}}, \quad \text{and} \quad \beta = \frac{1}{e^2 {M}^{4}_{\text{BI}}} = \frac{{g}^{2}}{{\alpha}^{4} {M}^{2}_{\text{pl}}},
\end{equation}
\noindent by defining ${\alpha} = \frac{{M}_{\text{BI}}}{{M}_{\text{pl}}}$. This model can be derived from DBI action of $R^2$ model in new minimal SUGRA, which is dual to DBI action of a massive vector multiplet. In the following two subsections, we are going to show this. 

\subsection{From DBI action of a massive vector multiplet, \dots}
\noindent The conformal SUGRA action is \cite{{1504.01221}, {1505.02235}}
\begin{equation}
\label{conformal SUGRA of DBI action}
{S} = \left[ \frac{1}{2} \left| {S}_{0} \right|^{2} \Phi \left( {\Lambda} + \bar{\Lambda} + g' V \right) \right]_{D} - \left[ h {S}_{0}^{3} X \right]_{F}
\end{equation}
\noindent where 
\begin{equation}
{S}^{3}_{0} X = \mathcal{W}^{\alpha} \mathcal{W}_{\alpha} - X {\Sigma}_{c} \left( Q \left( {\Lambda}, \bar{\Lambda} \right) \bar{S}_{0} \bar{X} \right), \\
\end{equation}
\noindent where $X$ is a chiral multiplet, ${\Sigma}_{c} \left( . \right)$ is the chiral projection operator in conformal SUGRA, $h$ is a real constant and $Q \left( {\Lambda}, \bar{\Lambda} \right)$ is a real function of $\Lambda$ and $\bar{\Lambda}$. After the super-conformal gauge fixings, integrating out all the auxiliary fields and choosing the $U \left( 1 \right)$ gauge condition $\theta = 0$ and taking $h = 1/4$ and $4 M^4 {\omega} = \text{exp} \left( 4J \left( C \right) / 3 \right)$, where $M$ is a positive constant, we have the bosonic part of Lagrangian of Eq.(\ref{conformal SUGRA of DBI action})

\begin{equation}
\begin{split}
\left. \mathcal{L} \right|_{B} =&\; \sqrt{-g} \left[  \frac{{M}_{\text{pl}}^{2}}{2} R - \frac{1}{2} J'' \left( C \right) {\partial}_{\mu} C {\partial}^{\mu} C - \frac{{g'}^{2}}{2} J'' \left( C \right) {A}_{\mu} {A}^{\mu} \right] \\
&\; + M^4 \left[ 1 - \sqrt{P} \sqrt{- \text{det} \left( {g}_{{\mu}{\nu}} + \frac{1}{M^4} \mathcal{F}_{{\mu}{\nu}} \right)} \right], \\
\end{split}
\end{equation}
\noindent where 
\begin{equation}
P = 1 + \frac{{g'}^{2}}{M^4} \left[ J \left( C \right) \right]^{2}. \\
\end{equation}

\subsection{From DBI action of $R^2$ model in new minimal SUGRA, \dots}

The DBI action of new minimal $\mathcal{N}=1$, $D=4$ SUGRA consists of a real linear compensator ${L}_{0}$, a real multiplet\footnote{The dual transformation is the same as that of Starobinsky case. } ${V}_{R} = \ln{\left( \frac{{L}_{0}}{|\mathcal{S}|^{2}} \right)}$ and a chiral multiplet $\mathcal{S}$. The action is given by \cite{{1504.01221}, {1505.02235}}
\begin{equation}
{S}_{\text{dual}} = \left[ \frac{3}{2} {L}_{0} {V}_{R} \right]_{D} + \left[ - \gamma X \left( \mathcal{W}, \bar{\mathcal{W}}, {L}_{0} \right) \right]_{F}, \\
\end{equation}
\noindent where $\gamma$ is a real constant, and $X \left( \mathcal{W}, \bar{\mathcal{W}}, {L}_{0} \right)$ is a solution of the equation
\begin{equation}
X = \mathcal{W}^{\alpha} \left( {V}_{R} \right) \mathcal{W}_{\alpha} \left( {V}_{R} \right) - \kappa {\Sigma}_{c} \left( |X|^{2} {L}^{-2}_{0} \right), \\
\end{equation}
\noindent where $\kappa$ is a positive constant. To solve $X$, we adopt the Lagrange multiplier multiplet $\mathcal{M}$ as 
\begin{equation}
\label{Action}
{S}_{\text{dual}} = \left[ \frac{3}{2} {L}_{0} {V}_{R} \right]_{D} + \left[ - \gamma X \right]_{F} + \left[ 2 \mathcal{M} \left( \mathcal{W}^{2} \left( {V}_{R} \right) - \kappa {\Sigma}_{c} \left( |X|^{2} {L}^{-2}_{0} \right) - X \right) \right]_{F}. \\
\end{equation}
\noindent After the gauge fixing of ${V}_{R}$ and the super-conformal symmetry, the bosonic part of the Lagrangian of Eq.(\ref{Action}) is given by
\begin{equation}
\begin{split}
\left. \mathcal{L} \right|_{B} =\;& \sqrt{-g} \left[ \frac{  {M}^{2}_{\text{pl}} }{2} R + \frac{3}{4} {B}_{\mu} {B}^{\mu} - \frac{3}{2} {A}^{\left( R \right)}_{\mu} {B}^{\mu} + 4 \kappa \left( \text{Re} \mathcal{M} \right) |{F}^{X}|^{2} - 4 \left( \text{Re} \mathcal{M} \right) {D}^{2}_{\left( R \right)} + 2 \left( \text{Re} \mathcal{M} \right) \mathcal{F}^{\left( R \right)}_{{\mu}{\nu}} \mathcal{F}^{\left( R \right) {\mu}{\nu}} \right. \\
&\left. - 2 i \left( \text{Im} \mathcal{M} \right) \mathcal{F}^{\left( R \right)}_{{\mu}{\nu}} \tilde{\mathcal{F}}^{\left( R \right) {\mu}{\nu}} - \left( \gamma + 2 \mathcal{M} \right) {F}^{X} - \left( \gamma + 2 \bar{\mathcal{M}} \right) \bar{F}^{\bar{X}} \right], \\
\end{split}
\end{equation}
\noindent where ${F}^{X}$ is the $F$ - term of $X$, ${D}_{\left( R \right)} := \frac{1}{3} \left(  {M}^{2}_{\text{pl}} R + \frac{3}{2} {B}_{\mu} {B}^{\mu} \right)$, ${A}^{\left( R \right)}_{\mu}$ is the vector component of ${V}_{R}$, ${B}_{\mu}$ is an auxiliary vector component of ${L}_{0}$, and $\mathcal{F}^{\left( R \right)}_{{\mu}{\nu}} = {\partial}_{\mu} {A}^{\left( R \right)}_{\nu} - {\partial}_{\nu} {A}^{\left( R \right)}_{\mu}$. After finding the equations of motion (E.O.M.)s of ${F}^{X}$, $\text{Re} \mathcal{M}$ and $\text{Im} \mathcal{M}$ and integrating out all of them, we obtain the on-shell action as follows

\begin{equation}
\label{On-shell Bosonic Lagrangian}
\begin{split}
\left. \mathcal{L} \right|_{B} =\;& \sqrt{-g} \left[ \frac{ {M}^{2}_{\text{pl}} }{2} R + \frac{3}{4} {B}_{\mu} {B}^{\mu} - \frac{3}{2} {A}^{\left( R \right)}_{\mu} {B}^{\mu} - \frac{\gamma}{\kappa} + \frac{\gamma}{\kappa} \sqrt{4 \kappa {D}^{2}_{\left( R \right)} - \text{det} \left( {\eta}_{ab} + \sqrt{\kappa} \mathcal{F}_{ab} \right)} \right]. \\
\end{split}
\end{equation}
\noindent The Lagrangian Eq.(\ref{On-shell Bosonic Lagrangian}) has the DBI structure to the vector ${A}^{\left( R \right)}_{\mu}$. But surprisingly, it also includes the curvature scalar inside the square root, which contributes to the high order correction. By taking ${A}^{\left( R \right)}_{\mu} = {B}_{\mu} = 0$, we obtain 
\begin{equation}
\left. \mathcal{L} \right|_{B} = \sqrt{-g} \frac{ {M}^{2}_{\text{pl}} }{2} \left[ R + \frac{2 g^2}{3 \beta} \left( \sqrt{1 + 12 \beta R^2} -1 \right) \right],  \\
\end{equation}

\noindent which gives BI-extended $F \left( R \right)$ term ${F}_{\text{BI}} \left( R \right) = R + \frac{2 g^2}{3 \beta} \left( \sqrt{1 + 12 \beta R^2} -1 \right)$. \cite{1505.02235} assumes the lowest scalar component of a vector multiplet is the inflaton field, while \cite{1807.08394} assumes the inflaton field comes from scalar curvature via tensor scalar transformation. This paper assumes the inflaton field coming from scalar curvature instead of the lowest scalar component of a vector multiplet. 

\subsection{BI-extended Potential}
To find the relationship between $R$ and $\phi$, by $F'(R) = {e}^{\sqrt{\frac{2}{3}} \frac{\phi}{{M}_{\text{pl}}}}$, we have

\begin{equation}
{F}'_{\text{BI}} \left( R \right) = 1+ \frac{8 {g}^{2} R}{\sqrt{1 + 12 \beta R^2}} = {e}^{\sqrt{\frac{2}{3}} \frac{\phi}{{M}_{\text{pl}}}} \\
\end{equation}
\noindent and on simplification we obtain
\begin{equation}
\label{R square}
R^2 = \frac{\left( {e}^{\sqrt{\frac{2}{3}} \frac{\phi}{{M}_{\text{pl}}}} - 1 \right)^2}{64 {g}^{4} - 12 \beta \left( {e}^{\sqrt{\frac{2}{3}} \frac{\phi}{{M}_{\text{pl}}}} - 1 \right)^2} \quad \Rightarrow \quad R = \frac{\pm \left( {e}^{\sqrt{\frac{2}{3}} \frac{\phi}{{M}_{\text{pl}}}} - 1 \right) }{\sqrt{ 64 {g}^{4} - 12 \beta \left( {e}^{\sqrt{\frac{2}{3}} \frac{\phi}{{M}_{\text{pl}}}} - 1 \right)^2 }}, \\
\end{equation}
\noindent Since $R$ is real, $R^2 \geq 0$, which implies\footnote{Note that $\beta > 0$. So, the inequality sign remains unchanged. }

\begin{equation}
\label{Real curvature criterion}
\frac{16 g^4}{3 \beta} \geq \left( {e}^{\sqrt{\frac{2}{3}} \frac{\phi}{{M}_{\text{pl}}}} - 1 \right)^{2} \quad \Rightarrow \quad 1 - \sqrt{\frac{16 g^4}{3 \beta}} \leq {e}^{\sqrt{\frac{2}{3}} \frac{\phi}{{M}_{\text{pl}}}} \leq 1 + \sqrt{\frac{16 g^4}{3 \beta}}. \\
\end{equation}

\noindent Also, from the physical point of view, since inflation occurred in dS spacetime, which has $R \geq 0$, and positive inflaton field value $\phi \geq 0$, and $R$ is a smooth function of $\phi$, it is physical to take
\begin{equation}
\label{Physical curvature}
R = \frac{ \left( {e}^{\sqrt{\frac{2}{3}} \frac{\phi}{{M}_{\text{pl}}}} - 1 \right) }{\sqrt{ 64 {g}^{4} - 12 \beta \left( {e}^{\sqrt{\frac{2}{3}} \frac{\phi}{{M}_{\text{pl}}}} - 1 \right)^2 }}, \\
\end{equation}
\noindent where the denominator satisfies the criterion stated in Eq.(\ref{Real curvature criterion}). Since $\chi = R$ in the Legendre transformation and the potential is given by

\begin{equation}
\label{Potential in chi}
{V}_{\text{BI}}(\chi) = \left( \frac{{M}^{2}_{\text{pl}}}{2} \right) \frac{{\chi}{F'(\chi)} - F(\chi)}{F'(\chi)^2} = \frac{g^2 \sqrt{12 \beta  \chi ^2+1} \left(\sqrt{12 \beta  \chi ^2+1}-1\right) M_{\text{pl}}^2}{3 \beta \left(\sqrt{12 \beta  \chi ^2+1}+8 g^2 \chi \right)^2}, \\
\end{equation}

\noindent substituting Eq.(\ref{Physical curvature}) into Eq.(\ref{Potential in chi}), we obtain the BI-extended potential in terms of the inflaton field $\phi$ as

\begin{equation}
\label{BI-extended Potential in phi}
{V}_{\text{BI}}(\phi) = \frac{ {M}_{\text{pl}}^2}{12 {\beta}} {e}^{-2 \sqrt{\frac{2}{3}} \frac{\phi}{ {M}_{\text{pl}} }} \left\{ 4 {g}^{2} - \sqrt{ 16{g}^{4} - 3 {\beta} \left( {e}^{\sqrt{\frac{2}{3}} \frac{\phi}{{M}_{\text{pl}}} } -1 \right)^{2} } \right\}. \\
\end{equation}

\section{Numerical Calculation}
\label{Numerical Calculation}
\subsection{Starobinsky model}

\begin{figure}[h!]
\centering
\includegraphics[width=72.5mm, height=55mm]{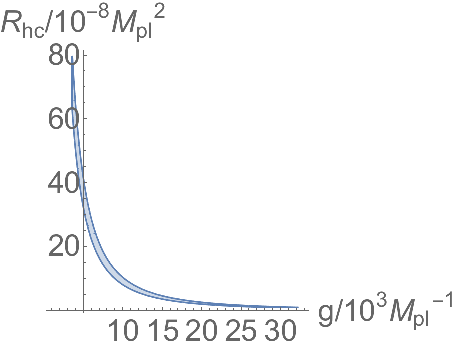} 
\includegraphics[width=72.5mm, height=55mm]{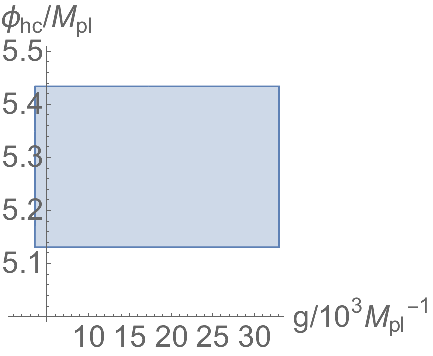} 
\caption{Left: $g$ and ${R}_{\text{hc}}$ based on the constraints listed in Table \ref{table:Planck data 2018 slow roll potential parameters and spectral indices}. Right: $g$ and ${\phi}_{\text{hc}}$ based on the constraints listed in Table \ref{table:Planck data 2018 slow roll potential parameters and spectral indices}. }
\label{fig: g and initial R and phi}
\end{figure}

\noindent Based on Planck 2018 data constraints \cite{Planck 2018 results. X. Constraints on inflation} listed in Table \ref{table:Planck data 2018 slow roll potential parameters and spectral indices}, we plot the region of possible values of $g$, scalar curvature at the first horizon crossing ${R}_{\text{hc}}/ {M}^{2}_{\text{pl}}$ and initial inflaton field value\footnote{ In this paper, the beginning of inflation is assumed to be the first horizon crossing. } ${\phi}_{\text{hc}}/ {M}_{\text{pl}}$ as shown in Figure \ref{fig: g and initial R and phi}. We can see the scalar curvature at the horizon crossing is roughly inversely proportional to $g$, even though its scale is about $O \left( 10^{-7} \right) {M}^{2}_{\text{pl}}$. The inflaton values at the first horizon crossing range from about $5.13 {M}_{\text{pl}}$ to $5.44 {M}_{\text{pl}}$ for all values of $g$ greater than $3500 {M}^{-1}_{\text{pl}}$. This range of $g$ implies that the mass scale $m$ given by $m^{-2} = 24 g^2$ has a range\footnote{Mathematically, it can be written as $-5.83212 \times 10^{-5} {M}_{\text{pl}} < m < 5.83212 \times 10^{-5} {M}_{\text{pl}}$. Given that $m$ is positive, we remove the negative side. } of $0<m< 5.83212 \times 10^{-5} {M}_{\text{pl}}$.

\subsection{BI-extended model}
\label{BI-extended model numerical calculation}

\begin{figure}[h!]
\centering
\includegraphics[width=145mm, height=120mm]{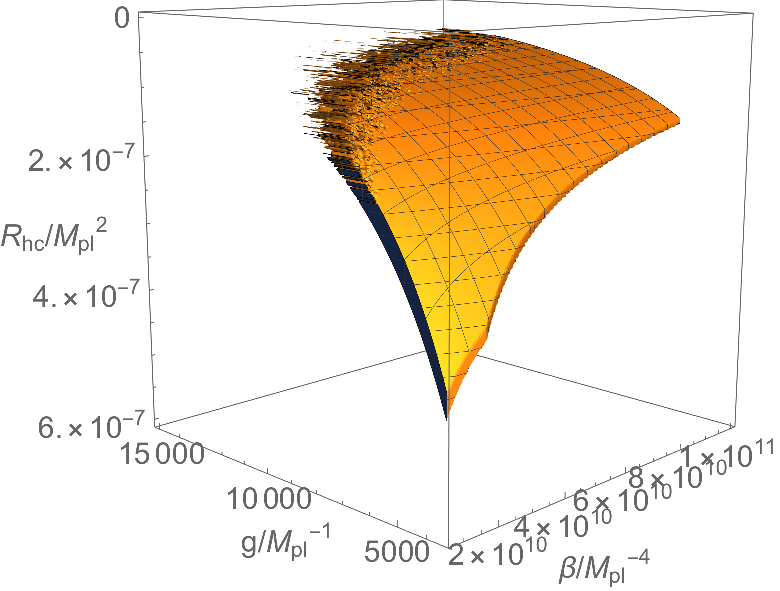} 
\caption{$g$, $\beta$ and ${R}_{\text{hc}}$ based on the constraints listed in Table \ref{table:Planck data 2018 slow roll potential parameters and spectral indices}. (Please pay attention to the magnitude direction of each axis. ) }
\label{fig: g, beta and initial R}
\end{figure}

\begin{figure}[h!]
\centering
\includegraphics[width=130mm, height=130mm]{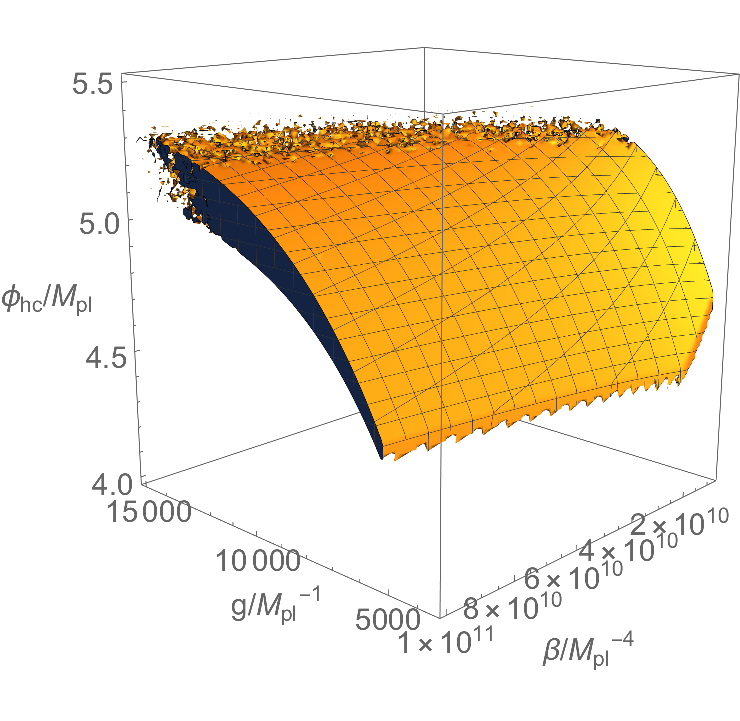} 
\caption{$g$, $\beta$ and ${\phi}_{\text{hc}}$ based on the constraints listed in Table \ref{table:Planck data 2018 slow roll potential parameters and spectral indices} for ${10}^{10} \leq {\beta} {M}_{\text{pl}}^{4} \leq 10^{11}$. (Please pay attention to the magnitude direction of each axis. ) }
\label{fig: g, beta and initial phi}
\end{figure}

\begin{figure}[h!]
\centering
\includegraphics[width=72.5mm, height=60mm]{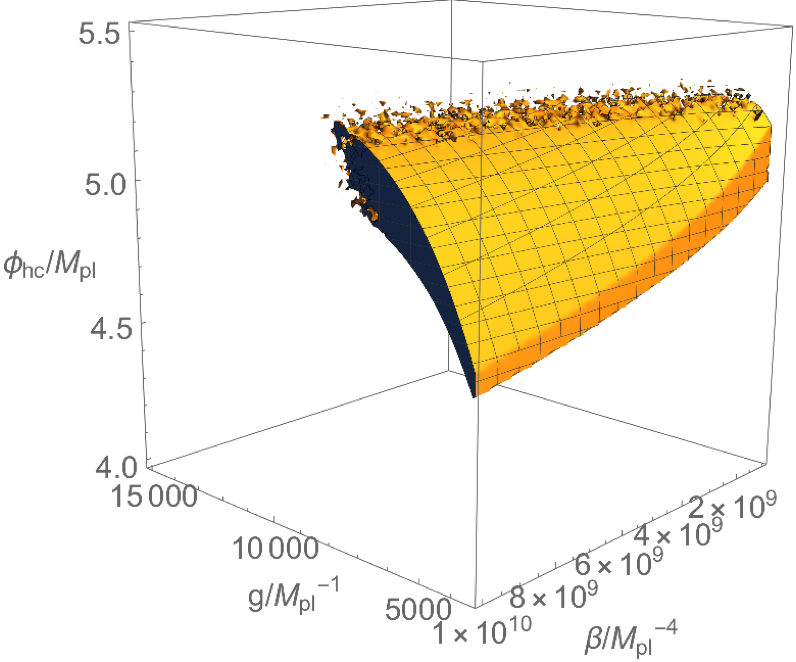} 
\includegraphics[width=72.5mm, height=60mm]{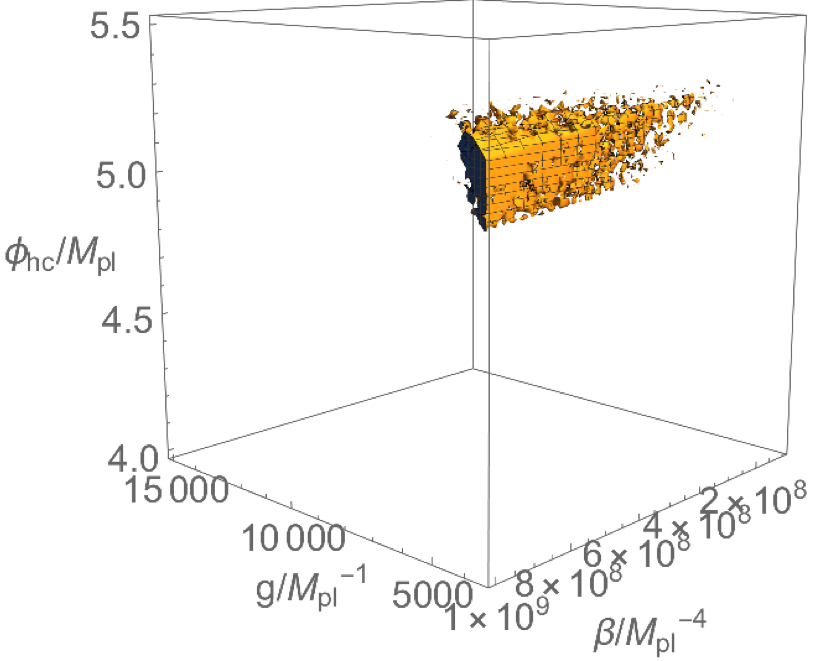} 
\caption{Left: $g$, $\beta$ and ${\phi}_{\text{hc}}$ based on the constraints listed in Table \ref{table:Planck data 2018 slow roll potential parameters and spectral indices} for ${10}^{9} \leq {\beta} {M}_{\text{pl}}^{4} \leq 10^{10}$. Right: Same for ${10}^{8} \leq {\beta} {M}_{\text{pl}}^{4} \leq 10^{9}$. (Please pay attention to the magnitude direction of each axis. ) }
\label{fig: g, small beta and initial phi}
\end{figure}

\noindent Based on Planck 2018 data constraints \cite{Planck 2018 results. X. Constraints on inflation} listed in Table \ref{table:Planck data 2018 slow roll potential parameters and spectral indices}, we plot the region of possible values of $g$, $\beta$, scalar curvature and inflaton field value at the horizon crossing ${\phi}_{\text{hc}}/ {M}_{\text{pl}}$ as shown in Figure \ref{fig: g, beta and initial R}, \ref{fig: g, beta and initial phi} and \ref{fig: g, small beta and initial phi}.\footnote{In previous work as shown in \cite{Cosmological Perturbations and Inflationary Theory}, inflaton values at the first horizon crossing are evaluated by obtaining the inflaton values at the end of inflation ${\phi}_{\text{end}}$ based on the condition of the end of inflation (${\epsilon}_{V} \left( {\phi}_{\text{end}} \right) = 1$ or ${\eta}_{V} \left( {\phi}_{\text{end}} \right) = 1$), and the slow-roll approximation of e-folding formula ${\Delta} N = \int^{{\phi}_{\text{end}}}_{{\phi}_{\text{hc}}} H dt$. But, in this paper, we find the inflaton values at the first horizon crossing based on the observation data as shown in Table \ref{table:Planck data 2018 slow roll potential parameters and spectral indices}, and then verify the consistency of e-folding number (should be between 50 and 60 e-folds) based on those obtained inflaton values at the first horizon crossing and the exact e-folding formula ${\Delta} N = \int^{{\phi}_{\text{end}}}_{{\phi}_{\text{hc}}} H dt$ as shown in Table \ref{table: Inflaton field evolution data}. } In fact, we can see the scalar curvature at the first horizon crossing is also roughly inversely proportional to $g$, even though its scale is about $O \left( 10^{-7} \right) {M}^{2}_{\text{pl}}$. The inflaton values at the first horizon crossing range from about $4.7 {M}_{\text{pl}}$ to $5.2 {M}_{\text{pl}}$ depending on some particular values of $g$ greater than $5000 {M}^{-1}_{\text{pl}}$, while $\beta$ is greater than $O \left( {10}^{8} \right) {M}^{-4}_{\text{pl}}$. Apart from this, we plot the possible values of $e$ and $\alpha$ in Fig. \ref{fig: e, alpha and initial phi} by the constraints in Table \ref{table:Planck data 2018 slow roll potential parameters and spectral indices} and Eq.(\ref{e and MBI}). The upper possible limits of $e$ and $\alpha$ are

\begin{equation}
e \leq 2.8 \times 10^{-4}, \quad \text{and} \quad \alpha \leq 0.41. 
\end{equation}

\begin{figure}[h!]
\centering
\includegraphics[width=72.5mm, height=65mm]{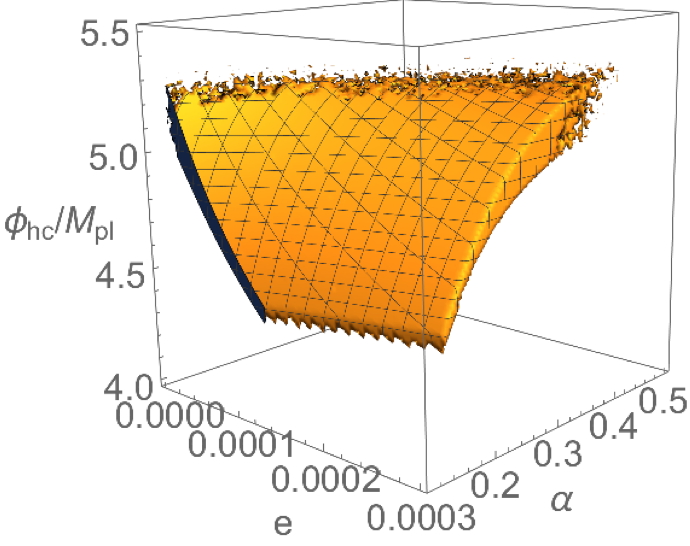} 
\includegraphics[width=72.5mm, height=67mm]{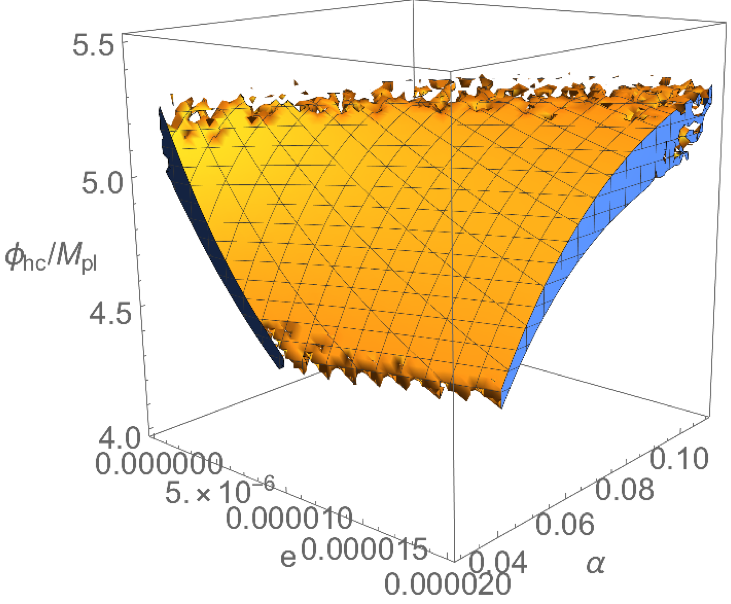} 
\caption{Left: $e$, $\alpha$ and ${\phi}_{\text{hc}}$ based on the constraints listed in Table \ref{table:Planck data 2018 slow roll potential parameters and spectral indices} for the ranges ${10}^{-5} \leq e \leq 3 \times {10}^{-4}$ and $0.11 \leq {\alpha} \leq 0.5$. Right: Same for the ranges ${10}^{-9} \leq e \leq 2 \times {10}^{-5}$ and $0.03 \leq {\alpha} \leq 0.11$. (Please pay attention to the magnitude direction and the scale of each axis. ) }
\label{fig: e, alpha and initial phi}
\end{figure}

\subsection{Potential graph under variations of $g$ and ${\beta}$}
\begin{figure}[h!]
\centering
\includegraphics[width=90mm, height=60mm]{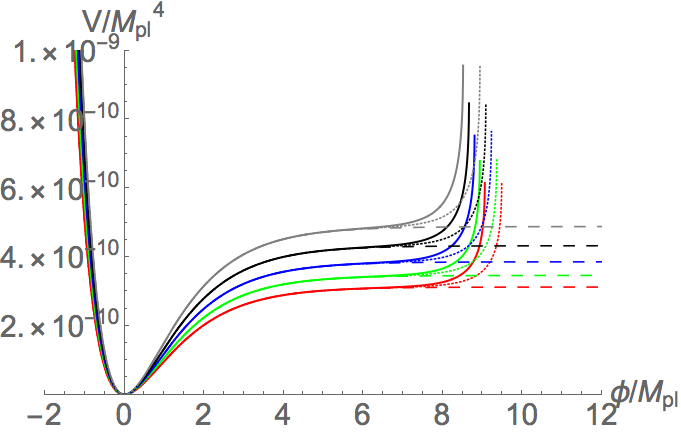} 
\includegraphics[width=55mm, height=35mm]{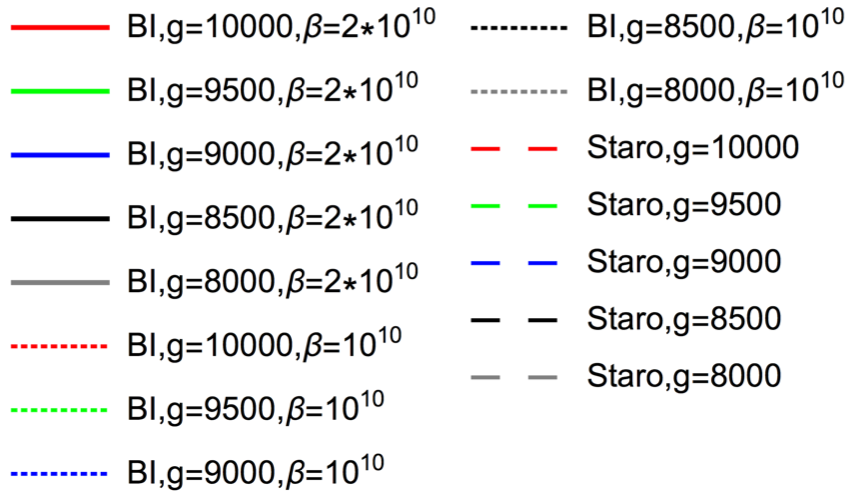} 
\caption{Left: This graph shows the potential against inflaton field values. The solid, tiny dashed and large dashed lines describe the BI-extended potential at $\beta = 2 \times 10^{10} {M}^{-4}_{\text{pl}}$, the counterpart at $\beta = 10^{10} {M}^{-4}_{\text{pl}}$ and the Starobinsky potential respectively. Around $\phi = 6 {M}_{\text{pl}}$, the potential lines are plotted from $g = 8000 {M}^{-1}_{\text{pl}}$ on the top to $g = 10000 {M}^{-1}_{\text{pl}}$ on the bottom. ) Since the universe started the inflation from about $5.1 - 5.4 {M}_{\text{pl}}$ as shown in Figure \ref{fig: g and initial R and phi} and Figure \ref{fig: g, beta and initial phi}, the universe has a higher initial potential scale as $g$ decreases. The universe rolled down to the trough at some inflaton value(s) such that either slow roll parameters $\epsilon$ or $\eta$ became close to $1$ as the end of inflation. Right: The plot legend. "BI" means BI-extended model while "Staro" means the Starobinsky model. }
\label{fig: Potential Plot}
\end{figure}

\noindent In this subsection, we are going to discuss the shapes of the BI-extended potential and the fingerprints in the ${n}_{s} - r$ graph when $g$ and $\beta$ vary. First, we can see that even though BI-extended potential Eq. (\ref{BI-extended Potential in phi}) has a similar shape with that of Starobinsky potential Eq. (\ref{Starobinsky potential}) during the inflation as shown in Fig.(\ref{fig: Potential Plot}). BI-extended potential has a higher value to start the inflation as $g$ decreases. It also has the "tail" at a larger inflaton field value as ${\beta}$ decreases. The tip of each tail ceases according to the real curvature criterion Eq. (\ref{Real curvature criterion}). \\

\subsection{${n}_{s} - r$ graph under variations of $g$ and ${\beta}$}
\begin{figure}[h!]
\begin{center}
\includegraphics[width=70mm, height=55mm]{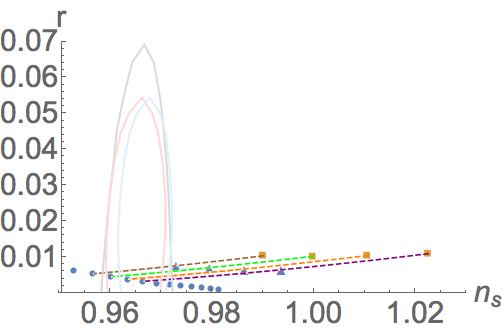} 
\includegraphics[width=70mm, height=55mm]{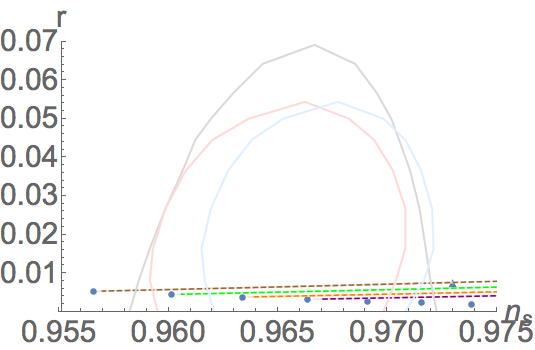} 
\caption{Left: The tensor-to-scalar ratio $r$ against the scalar spectral index ${n}_{s}$ are plotted as above. The light color lines represent the boundaries of the observation data of 68 \% CL in Planck 2018 (Light gray for TT,TE,EE+lowE+lensing, light red for TT,TE,EE+lowE+lensing+BK14 and light blue for TT,TE,EE+lowE+lensing+BK14+BAO). The blue dots represent the fingerprints of Starobinsky model, from the inflaton values at the first horizon crossing ${\phi}_{\text{hc}} = 5 \; {M}_{\text{pl}}$ on the left to ${\phi}_{\text{hc}} = 6.1 \; {M}_{\text{pl}}$ on the right with a spacing of $0.1 \; {M}_{\text{pl}}$. The square points represent the fingerprints of BI-extended model at $g = 5000 \; {M}^{-1}_{\text{pl}}$ and $\beta = 2 \times {10}^{10} \; {M}^{-4}_{\text{pl}}$ while the triangle points represent the counterparts at $g = 5000 \; {M}^{-1}_{\text{pl}}$ and $\beta = {10}^{10} \; {M}^{-4}_{\text{pl}}$. The brown, green, orange and purple dashed lines represent ${\phi}_{\text{hc}} = 5.1 \; {M}_{\text{pl}}, \; 5.2 \; {M}_{\text{pl}}, \; 5.3 \; {M}_{\text{pl}}, \; 5.4 \; {M}_{\text{pl}}$ respectively. As $g$ increases from $5000 \; {M}^{-1}_{\text{pl}}$ to $15000 \; {M}^{-1}_{\text{pl}}$, they go along the corresponding dashed lines towards the blue points (Starobinsky fingerprints) from the corresponding square and triangle points respectively. Right: This graph focuses on the inner observation regions.} 
\label{fig: ns-r graph}
\end{center}
\end{figure}

\noindent Variations of $g$ and $\beta$ give various fingerprints in the graph of the tensor-to-scalar ratio $r$ against the scalar spectral index ${n}_{s}$ as shown in Fig.(\ref{fig: ns-r graph}). With the reference points of Starobinsky model shown as blue points, BI-extended model gives the fingerprints tending to those blue points as $g$ increases. BI-extended fingerprints start from $\beta = 10^{10} {M}^{-4}_{\text{pl}}$ and $\beta = 2 \times 10^{10} {M}^{-4}_{\text{pl}}$ at triangle and square points respectively. This implies that BI-extended fingerprints are away from the observation data as ${\beta}$ increases.

\section{Evolutions of inflaton field}
\label{Evolutions of inflaton field}

\begin{figure}[h!]
\begin{center}
\includegraphics[width=65mm, height=55mm]{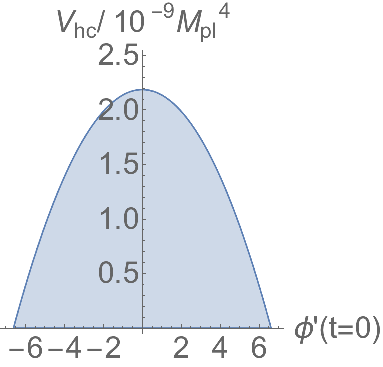} 
\includegraphics[width=80mm, height=55mm]{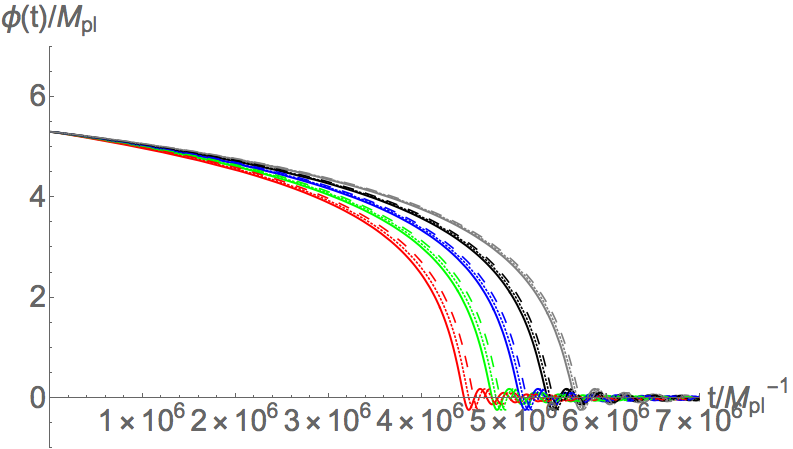} 
\caption{Left: These region plots show the potential scale against the speed of inflaton field at the first horizon crossing based on Table \ref{table:Planck data 2018 slow roll potential parameters and spectral indices}. The horizontal axis ($x$-axis) and the vertical axis ($y$-axis) represent the speed of inflaton field $\dot{\phi} \left( t \right) / {10}^{-5} {M}^{2}_{\text{pl}}$ and the potential scale $V \left( {\phi} \right)/ {10}^{-9} {M}^{4}_{\text{pl}}$ (at the first horizon crossing) respectively. Right: This graph shows the time evolution of inflaton field with respect to cosmic time $t$. The colors can be referred to the plot legend in Fig.(\ref{fig: Potential Plot}). }
\label{fig: Initial Inflaton Speed Values}
\end{center}
\end{figure}

\begin{figure}[h!]
\begin{center}
\includegraphics[width=150mm, height=90mm]{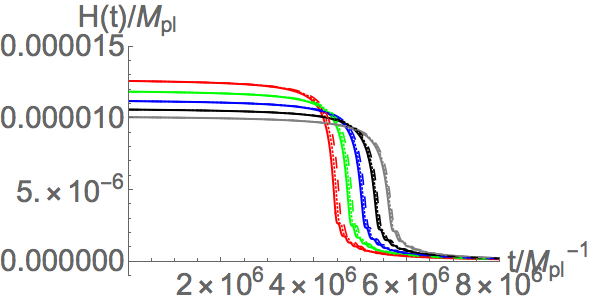} 
\caption{This graph shows the time evolution of the Hubble parameter with respect to cosmic time $t$. The colors can be referred to the plot legend in Fig.(\ref{fig: Potential Plot}). }
\label{fig: Hubble Evolution Time}
\end{center}
\end{figure}

\begin{table}[h!]
\begin{center}
\begin{tabular}{ |c|c|c|c|c|c| }
\hline
$\text{Color, style}$ & $\text{Model}$ & $g/{10}^{3} {M}^{-1}_{\text{pl}}$ & ${\beta}/{10}^{10} {M}^{-4}_{\text{pl}}$ & $t/{M}^{-1}_{\text{pl}} (N=50)$ & $t/{M}^{-1}_{\text{pl}} (N=60)$ \\
\hline 
$\text{Red, Thick}$ & $\text{BI}$ & $10$ & $2$ & $5.069835 \times {10}^{6}$ & $>3.985 \times {10}^{7}$ \\
\hline
$\text{Green, Thick}$ & $\text{BI}$ & $9.5$ & $2$ & $4.817745 \times {10}^{6}$ & $> 5.098 \times {10}^{7}$ \\
\hline
$\text{Blue, Thick}$ & $\text{BI}$ & $9$ & $2$ & $4.565945  \times {10}^{6}$ & $> 5.708 \times {10}^{7}$ \\
\hline
$\text{Black, Thick}$ & $\text{BI}$ & $8.5$ & $2$ & $4.314575 \times {10}^{6}$ & $> 6.314 \times {10}^{7}$ \\
\hline
$\text{Gray, Thick}$ & $\text{BI}$ & $8$ & $2$ & $4.063815 \times {10}^{6}$ & $> 6.726 \times {10}^{7}$ \\
\hline 
$\text{Red, Dashed(tiny)}$ & $\text{BI}$ & $10$ & $1$ & $5.0667135 \times {10}^{6}$ & $> 5.052 \times {10}^{7}$ \\
\hline
$\text{Green, Dashed(tiny)}$ & $\text{BI}$ & $9.5$ & $1$ & $4.814035 \times {10}^{6}$ & $> 6.094 \times {10}^{7}$ \\
\hline
$\text{Blue, Dashed(tiny)}$ & $\text{BI}$ & $9$ & $1$ & $4.561495 \times {10}^{6}$ & $> 7.117 \times {10}^{7}$ \\
\hline
$\text{Black, Dashed(tiny)}$ & $\text{BI}$ & $8.5$ & $1$ & $4.309125 \times {10}^{6}$ & $> 8.233 \times {10}^{7}$ \\
\hline
$\text{Gray, Dashed(tiny)}$ & $\text{BI}$ & $8$ & $1$ & $4.057005 \times {10}^{6}$ & $> 9.216 \times {10}^{7}$ \\
\hline 
$\text{Red, Dashed(long)}$ & $\text{Staro}$ & $10$ & $\text{N/A}$ & $5.06373 \times {10}^{6}$ & $> 3.962  \times {10}^{7}$ \\
\hline
$\text{Green, Dashed(long)}$ & $\text{Staro}$ & $9.5$ & $\text{N/A}$ & $4.810548 \times {10}^{6}$ & $> 4.933 \times {10}^{7}$ \\
\hline
$\text{Blue, Dashed(long)}$ & $\text{Staro}$ & $9$ & $\text{N/A}$ & $4.557355 \times {10}^{6}$ & $> 5.273 \times {10}^{7}$ \\
\hline
$\text{Black, Dashed(long)}$ & $\text{Staro}$ & $8.5$ & $\text{N/A}$ & $4.304175 \times {10}^{6}$ & $> 5.422 \times {10}^{7}$ \\
\hline
$\text{Gray, Dashed(long)}$ & $\text{Staro}$ & $8$ & $\text{N/A}$ & $4.050985 \times {10}^{6}$ & $> 5.934 \times {10}^{7}$ \\
\hline
\end{tabular}
\end{center}
\caption{Inflaton field evolution data starting from ${\phi} = 5.3 {M}_{\text{pl}}$ and $\dot{\phi} (t=0) = 0$. The pairs of color and style can be referred to the plot legend in Fig.(\ref{fig: Potential Plot}). }
\label{table: Inflaton field evolution data}
\end{table}

\noindent In this part, we are going to investigate how the inflaton field evolves as $\left( g, \beta \right)$ vary. First of all, we are going to find the initial rate of change of inflaton field $\dot{\phi} \left( t \right)$. By adopting the Friedman equation in Eq.(\ref{Einstein equations}) and the constraints in Table \ref{table:Planck data 2018 slow roll potential parameters and spectral indices} as demonstrated in \cite{Initial conditions for Inflation in an FRW Universe}, we obtain the region plot in Fig.(\ref{fig: Initial Inflaton Speed Values}). We can see that the region is enclosed by a parabolic curve with the x-axis since the speed of inflaton field $\dot{\phi} \left( t \right)$ ($x$-axis) is parabolic in the Friedman equation while the potential ($y$-axis) is linear. We can also understand that $\dot{\phi} \left( {t=0} \right)$ ranges from $-6 \times {10}^{-5} {M}^{2}_{\text{pl}}$ to $6 \times {10}^{-5} {M}^{2}_{\text{pl}}$ while ${V}_{\text{hc}}$ ranges from $0$ to $2.2 \times {10}^{-9} {M}^{4}_{\text{pl}}$. \\

\noindent Apart from this, we plot time evolutions of ${\phi} \left( {t} \right)/ {M}_{\text{pl}}$ and $H \left( t \right)/ {M}_{\text{pl}}$ with initial conditions ${\phi} \left( {t=0} \right) = 5.3 {M}_{\text{pl}}$ and $\dot{\phi} \left( {t=0} \right) = 0$ under variations of $g$ and ${\beta}$ in Fig.(\ref{fig: Initial Inflaton Speed Values}) and Fig.(\ref{fig: Hubble Evolution Time}) respectively. In Fig.(\ref{fig: Initial Inflaton Speed Values}), one can see that all colors of ${\phi} \left( t \right)$ decay continuously from $5.3 {M}_{\text{pl}}$ to $0.5 {M}_{\text{pl}}$ and then oscillate around $0$. The evolutions carry out oscillations earlier for larger $g$ and $\beta$. On the other hand, in Fig.(\ref{fig: Hubble Evolution Time}), one can see that even though different colors start from various initial Hubble values, all colors of $H \left( t \right)$ remain constant from $t=0$ to $t=4 \times {10}^{6} {M}^{-1}_{\text{pl}}$, and then start to decay to approach $0$ between $t=4 \times {10}^{6} {M}^{-1}_{\text{pl}}$ and $t=6 \times {10}^{6} {M}^{-1}_{\text{pl}}$. This verifies that the initial stage of inflation can be approximated as "quasi-static"\footnote{The words "quasi static" means the Hubble parameter $H \left( t \right)$ is constant. }. The evolutions decay earlier for larger $g$ and $\beta$ as well.

\section{Discussions}
\label{Discussions}
\noindent Basically, from the potential graph shown in Fig.(\ref{fig: Potential Plot}), we can understand that the inflation processes described by BI-extended and Starobinsky models are similar, even though their shapes are different in the regions out of inflation dynamics. We can also know that the feasible scales of BI-extended model are $g \approx O \left( {10}^{3} \right) {M}^{-1}_{\text{pl}}$ and $\beta \approx O \left( {10}^{8} \right) {M}^{-4}_{\text{pl}}$. Since the scale of the scalar curvature at the first horizon crossing is given by $R \approx O \left( {10}^{-7} \right) {M}^{2}_{\text{pl}}$, by Eq. (\ref{BI-extended model}), we know that the total scale of $12 {\beta} {R}^{2}$ becomes about $O \left( {10}^{-4} \right)$, which is much less than $1$. The square root function in Eq. (\ref{BI-extended model}) can be well approximated as $\sqrt{1+12 {\beta} {R}^{2}} \approx 1 + 6 {\beta} {R}^{2}$, which reduces BI-extended model to the Starobinsky counterpart. Since the inflaton value decreases from about $5.3 {M}_{\text{pl}}$ to about $1 {M}_{\text{pl}}$ during inflation process, by Eq.(\ref{R square}), the (positive) value of scalar curvature decreases, which keeps the approximation of the square root function valid. \\

\vspace{2mm}

\noindent Apart from this, in Fig.(\ref{fig: Potential Plot}), there are sharply increasing "tails" in BI-extended potentials as $\phi$ increases, and then stop at some certain values. In fact, as $\phi$ increases, ${e}^{\sqrt{\frac{2}{3}} \frac{\phi}{{M}_{\text{pl}}}}$ in Eq.(\ref{BI-extended Potential in phi}) also increases significantly, resulting in a decline of $16 {g}^{4} - 3 {\beta} \left( {e}^{\sqrt{\frac{2}{3}} \frac{\phi}{{M}_{\text{pl}}}} - 1 \right)^{2}$ from a positive value to $0$, at which the curvature criterion Eq.(\ref{Real curvature criterion}) takes effect. On the other side, as $\phi$ decreases to negative values through $0$, the exponential factor ${e}^{\sqrt{\frac{2}{3}} \frac{\phi}{{M}_{\text{pl}}}}$ declines and approaches to $0$, leading to a constant value for $16 {g}^{4} - 3 {\beta} \left( {e}^{\sqrt{\frac{2}{3}} \frac{\phi}{{M}_{\text{pl}}}} - 1 \right)^{2}$. However, the factor ${e}^{-2 \sqrt{\frac{2}{3}} \frac{\phi}{{M}_{\text{pl}}}}$ rises dramatically to approach infinity, causing surges of BI-extended potentials. \\

\vspace{2mm}

\noindent In addition, even though the variations of $\left( g, \beta \right)$ of BI-extended models on $ \left( {n}_{s}, r \right)$ are along their corresponding dashed lines, from Fig.(\ref{fig: ns-r graph}) we can understand that observations favor large $g$ and small $\beta$. BI-extended models also agree with the inflaton values at the first horizon crossing ranging from $5.1 {M}_{\text{pl}}$ to $5.4 {M}_{\text{pl}}$, which have the same value tolerance as that of Starobinsky potential at the first horizon crossing. Since BI-extended model gives nearly the same effective inflation dynamics as Starobinsky model, further investigation is required to verify which one, BI-extended or Starobinsky model, is the correct model for describing the beginning of inflation. It is also meaningful to find the theoretical reasons why observation data favor large $g$ and and small $\beta$ in BI-extended model. Nevertheless, BI-extended model can be one of the extensions of Starobinsky model for describing the inflation dynamics at the first horizon crossing, and the high order terms can refine the prediction of original Starobinsky model.

\section{Conclusions}
\label{Conclusions}
\noindent To conclude, we analyze BI-extended model in a complete form and compare the predictions with that of Starobinsky model. Under the observation constraints of Planck 2018, we find that the inflation processes described by BI-extended and Starobinsky models are nearly the same. The required scales of $g$ and $\beta$ are approximately at least $O \left( {10}^{3} \right) {M}^{-1}_{\text{pl}}$ and $O \left( {10}^{8} \right) {M}^{-4}_{\text{pl}}$ respectively, which give the upper limit of the coupling $e$ and the ratio of ${M}_{\text{BI}}$ to ${M}_{\text{pl}}$ as $e \leq 2.8 \times 10^{-4}$ and ${\alpha} \leq 0.41$ respectively. Furthermore, we investigate the changes of predictions of $\left( {n}_{s}, r \right)$ and the evolutions of inflaton field under the variations of $\left( g, {\beta} \right)$. BI-extended model can be one of the extensions of Starobinsky model for high energy scale. 

\section{Acknowledgements}
\noindent The author thanks Prof. Hiroyuki ABE very much for suggestion and useful discussions. The author is supported by Waseda University Scholarship. 

\appendix

\section{Appendix: Potential and its derivatives}
\noindent In this section, we are going to discuss the full derivation of the potential and its derivatives for slow-roll parameters. Given that the derivative of ${\chi}$ with respect to ${\phi}$ is given by
\begin{equation}
\frac{d {\chi}}{d {\phi}} = \frac{1}{{M}_{\text{pl}}} \sqrt{\frac{2}{3}} \frac{F' \left( {\chi} \right)}{F'' \left( {\chi} \right)}, \\
\end{equation}
\noindent its first derivative with respect to $\chi$ is given by
\begin{equation}
\frac{d}{d {\chi}} \left( \frac{d {\chi}}{d {\phi}} \right) = \frac{1}{{M}_{\text{pl}}} \sqrt{\frac{2}{3}} \frac{\left[ F''(\chi )^2-F^{(3)}(\chi ) F'(\chi) \right]}{F''(\chi )^2}. \\
\end{equation}
\noindent Its second derivative with respect to $\chi$ is given by
\begin{equation}
\frac{d^{2}}{d {\chi}^{2}} \left( \frac{d {\chi}}{d {\phi}} \right) = \frac{-1}{{M}_{\text{pl}}} \sqrt{\frac{2}{3}} \frac{ \left[ F^{(3)}(\chi ) F''(\chi )^2-2 F^{(3)}(\chi )^2 F'(\chi)+F^{(4)}(\chi ) F'(\chi ) F''(\chi )\right]}{F''(\chi )^3}. \\
\end{equation}
\noindent Its third derivative with respect to $\chi$ is given by
\begin{equation}
\begin{split}
\frac{d^{3}}{d {\chi}^{3}} \left( \frac{d {\chi}}{d {\phi}} \right) =& \; \frac{-1}{{M}_{\text{pl}}} \sqrt{\frac{2}{3}} \frac{1}{F''(\chi )^4} \left[ 2 F^{(4)}(\chi ) F''(\chi )^3+6 F^{(3)}(\chi )^3 F'(\chi) \right. \\
& \left. +\left( F^{(5)}(\chi ) F'(\chi )-3 F^{(3)}(\chi )^2\right) F''(\chi )^2 - 6 F^{(3)}(\chi ) F^{(4)}(\chi ) F'(\chi ) F''(\chi)\right], \\
\end{split}
\end{equation}
\noindent where $F' \left( {\chi} \right)$, $F'' \left( {\chi} \right)$, $F^{(3)} \left( {\chi} \right)$ and $F^{(4)} \left( {\chi} \right)$ are the first, second, third and fourth derivatives of $F \left( {\chi} \right)$ with respect to ${\chi}$ respectively. The potential is given by
\begin{equation}
V \left( {\chi} \right) = \left( \frac{{M}^{2}_{\text{pl}}}{2} \right) \frac{{\chi} F' \left( {\chi} \right) - F \left( {\chi} \right)}{F' \left( {\chi} \right)^{2}}, \\
\end{equation}
\noindent and its first, second and third derivatives with respect to ${\chi}$ are given by
\begin{equation}
V' \left( {\chi} \right) = \left( \frac{{M}^{2}_{\text{pl}}}{2} \right) F'' (\chi) \frac{2 F (\chi) - {\chi} F' (\chi)}{{F' (\chi)}^{3}}, \\
\end{equation}

\begin{equation}
\begin{split}
V'' \left( {\chi} \right) =& \frac{M_{\text{pl}}^2}{2 F'(\chi )^4} \left[ -6 F(\chi ) F''(\chi )^2+F'(\chi )^2 \left(F''(\chi )-\chi 
   F^{(3)}(\chi )\right) \right. \\
   & \left. +2 F'(\chi ) \left(F(\chi ) F^{(3)}(\chi )+\chi  F''(\chi)^2\right)\right], \\
\end{split}
\end{equation}

\begin{equation}
\begin{split}
V''' \left( {\chi} \right) =& \frac{- M_{\text{pl}}^2}{2 F'(\chi )^5} \left[ -24 F(\chi ) F''(\chi )^3+\left(\chi  F^{(4)}(\chi )-2
   F^{(3)}(\chi )\right) F'(\chi )^3 \right. \\ 
   & \left. +6 F'(\chi ) \left(\chi  F''(\chi )^3+3 F(\chi )
   F^{(3)}(\chi ) F''(\chi )\right) \right. \\ 
   & \left. +F'(\chi )^2 \left(-2 F(\chi ) F^{(4)}(\chi )+6
   F''(\chi )^2-6 \chi  F^{(3)}(\chi ) F''(\chi )\right)\right], \\
\end{split}
\end{equation}

\begin{equation}
\begin{split}
V'''' \left( {\chi} \right) =& \; \frac{- {M}_{\text{pl}}^2}{2 F'(\chi )^6} \left[ 120 F(\chi ) F''(\chi )^4+\left(\chi  F^{(5)}(\chi )-3 F^{(4)}(\chi )\right) F'(\chi )^4 \right. \\
& \left. -24 F'(\chi ) \left(\chi  F''(\chi )^4+6 F(\chi ) F^{(3)}(\chi ) F''(\chi )^2\right) \right. \\
& \left. -6 F'(\chi)^2 \left(-3 F(\chi ) F^{(3)}(\chi )^2+6 F''(\chi )^3-4 F(\chi ) F^{(4)}(\chi ) F''(\chi )-6 \chi  F^{(3)}(\chi ) F''(\chi)^2\right) \right. \\
& \left. -2 F'(\chi )^3 \left(F(\chi ) F^{(5)}(\chi )+3 \chi  F^{(3)}(\chi)^2+\left(4 \chi  F^{(4)}(\chi )-14 F^{(3)}(\chi )\right) F''(\chi )\right) \right]. \\
\end{split}
\end{equation}

\noindent The first, second, third and fourth derivatives of the potential with respect to ${\phi}$ are given by
\begin{equation}
\frac{d V}{d {\phi}} = V' \left( {\chi} \right) \frac{d {\chi}}{d {\phi}} = \frac{{M}_{\text{pl}}}{\sqrt{6}} \frac{2 F - {\chi} F'}{{F'}^{2}}, \\
\end{equation}

\begin{equation}
\begin{split}
\frac{d^{2} V}{d {\phi}^{2}} =& \; V'' \left( {\chi} \right) \left( \frac{d {\chi}}{d {\phi}} \right)^{2} + V' \left( {\chi} \right) \frac{d {\chi}}{d {\phi}} \frac{d}{d {\chi}} \left( \frac{d {\chi}}{d {\phi}} \right) \\
=& \; \frac{1}{3} \left[ \frac{1}{F''(\chi )}+\frac{\chi }{F'(\chi )}-\frac{4 F(\chi )}{F'(\chi)^2}\right],  \\
\end{split}
\end{equation}

\begin{equation}
\begin{split}
\frac{d^{3} V}{d {\phi}^{3}} =& \; {V}^{\left( 3 \right)} \left( {\chi} \right) \left( \frac{d {\chi}}{d {\phi}} \right)^{3} + 3 {V}'' \left( {\chi} \right) \left( \frac{d {\chi}}{d {\phi}} \right)^{2} \frac{d}{d {\chi}} \left( \frac{d {\chi}}{d {\phi}} \right) + V' \left( {\chi} \right) \frac{d {\chi}}{d {\phi}} \left[ \frac{d}{d {\chi}} \left( \frac{d {\chi}}{d {\phi}} \right) \right]^{2} \\
&+ V' \left( {\chi} \right) \left( \frac{d {\chi}}{d {\phi}} \right)^{2} \frac{d^{2}}{d {\chi}^{2}} \left( \frac{d {\chi}}{d {\phi}} \right) \\
=&\; \frac{-1}{3 {M}_{\text{pl}}} \sqrt{\frac{2}{3}} \frac{1}{F'(\chi)^2 F''(\chi )^3}  \left[ - 8 F(\chi ) F''(\chi )^3+F^{(3)}(\chi ) F'(\chi )^3 + 3F'(\chi )^2 F''(\chi )^2 \right. \\
&\; \left. + \chi  F'(\chi ) F''(\chi )^3 \right] \\
\end{split}
\end{equation}

\begin{equation}
\begin{split}
\frac{d^{4} V}{d {\phi}^{4}} =& \; {V}^{\left( 4 \right)} \left( {\chi} \right) \left( \frac{d {\chi}}{d {\phi}} \right)^{4} + 6 {V}^{\left( 3 \right)} \left( {\chi} \right) \left( \frac{d {\chi}}{d {\phi}} \right)^{3} \frac{d}{d {\chi}} \left( \frac{d {\chi}}{d {\phi}} \right) + 7 {V}^{\left( 2 \right)} \left( {\chi} \right) \left( \frac{d {\chi}}{d {\phi}} \right)^{2} \left[ \frac{d}{d {\chi}} \left( \frac{d {\chi}}{d {\phi}} \right) \right]^{2} \\
& \; + 4 {V}^{\left( 2 \right)} \left( {\chi} \right) \left( \frac{d {\chi}}{d {\phi}} \right)^{3} \frac{d^2}{d {\chi}^2} \left( \frac{d {\chi}}{d {\phi}} \right) + 4 {V}' \left( {\chi} \right) \left( \frac{d {\chi}}{d {\phi}} \right)^{2} \left[ \frac{d}{d {\chi}} \left( \frac{d {\chi}}{d {\phi}} \right) \right] \frac{d^2}{d {\chi}^2} \left( \frac{d {\chi}}{d {\phi}} \right) \\
& \; + {V}' \left( {\chi} \right) \left( \frac{d {\chi}}{d {\phi}} \right)^{3} \frac{d^3}{d {\chi}^3} \left( \frac{d {\chi}}{d {\phi}} \right) \\
=& \frac{-2}{9 {M}^{2}_{\text{pl}} F'(\chi )^2 F''(\chi)^6} \left[ 18 F(\chi ) F''(\chi )^6-2 F'(\chi ) \left(\chi  F''(\chi )^6+3 F(\chi )
   F^{(3)}(\chi ) F''(\chi )^4\right) \right. \\
  & \left. -F^{(3)}(\chi ) F'(\chi )^3 \left(2 F(\chi) F^{(3)}(\chi)^2 + 2 F''(\chi )^3+3 \chi  F^{(3)}(\chi ) F''(\chi )^2\right) \right. \\
  & \left. + F'(\chi)^2 \left(-7 F''(\chi )^5+3 \chi  F^{(3)}(\chi ) F''(\chi )^4+6 F(\chi ) F^{(3)}(\chi)^2 F''(\chi )^2\right) \right. \\
   & \left. +F'(\chi )^4 \left(\chi  F^{(3)}(\chi )^3+F^{(4)}(\chi ) F''(\chi )^2-3 F^{(3)}(\chi )^2 F''(\chi )\right)\right], \\
\end{split}
\end{equation}

\noindent and the normalized derivatives are given by
\begin{equation}
\frac{1}{V} \frac{d V}{d {\phi}} = - \sqrt{\frac{2}{3}} \frac{1}{{M}_{\text{pl}}} \frac{{\chi} F' - 2 F}{{\chi} F' - F}. \\
\end{equation}

\begin{equation}
\frac{1}{V} \frac{d^{2} V}{d {\phi}^{2}} = -\frac{2 \left[ -4 F(\chi ) F''(\chi )+F'(\chi )^2+\chi  F'(\chi ) F''(\chi )\right]}{3 {M}^{2}_{\text{pl}} \left[ F(\chi)-\chi  F'(\chi )\right] F''(\chi)}. \\
\end{equation}

\begin{equation}
\begin{split}
\frac{1}{V} \frac{d^{3} V}{d {\phi}^{3}} =& \; \frac{1}{{M}^{3}_{\text{pl}}} \left( \frac{2}{3} \right)^{\frac{3}{2}} \frac{ \left[ -8 F(\chi ) F''(\chi )^3+F^{(3)}(\chi ) F'(\chi )^3 + 3 F'(\chi )^2 F''(\chi )^2+\chi  F'(\chi ) F''(\chi )^3\right]}{ \left[ F(\chi)-\chi  F'(\chi) \right] F''(\chi )^3}, \\
\end{split}
\end{equation}

\begin{equation}
\begin{split}
\frac{1}{V} \frac{d^{4} V}{d {\phi}^{4}} =& \; \frac{4}{9 {M}^{4}_{\text{pl}} \left(F(\chi )-\chi 
   F'(\chi )\right) F''(\chi )^6} \left[18 F(\chi ) F''(\chi )^6 \right. \\
   & \left. -2 F'(\chi ) \left(\chi  F''(\chi )^6+3 F(\chi) F^{(3)}(\chi ) F''(\chi )^4\right) \right. \\
   & \; \left. -F^{(3)}(\chi ) F'(\chi )^3 \left(2 F(\chi ) F^{(3)}(\chi )^2+2 F''(\chi )^3+3 \chi  F^{(3)}(\chi ) F''(\chi )^2\right) \right. \\
   & \left. +F'(\chi)^2 \left(-7 F''(\chi )^5+3 \chi  F^{(3)}(\chi ) F''(\chi )^4+6 F(\chi ) F^{(3)}(\chi)^2 F''(\chi )^2\right) \right. \\
   & \left. +F'(\chi )^4 \left(\chi  F^{(3)}(\chi )^3+F^{(4)}(\chi) F''(\chi )^2-3 F^{(3)}(\chi )^2 F''(\chi )\right)\right]. \\
\end{split}
\end{equation}

\noindent Based on the above derivatives, the slow roll parameters defined as
\begin{equation}
\begin{split}
{\epsilon}_{V} &\equiv \text{1st order} \equiv \frac{1}{2}{M}_{\text{pl}}^{2} \left( \frac{{V}' \left( {\phi} \right)}{V \left( {\phi} \right)} \right)^{2}, \\
{\eta}_{V} &\equiv \text{2nd order} \equiv {M}_{\text{pl}}^{2} \left( \frac{{V}'' \left( {\phi} \right)}{V \left( {\phi} \right)} \right), \\
{\xi}_{V} &\equiv \text{3rd order} \equiv {M}_{\text{pl}}^{4} \left( \frac{{V}' \left( {\phi} \right) {V}''' \left( {\phi} \right)}{{V \left( {\phi} \right)}^{2}} \right), \\
{\omega}_{V} &\equiv \text{4th order} \equiv {M}^{6}_{\text{pl}} \left( \frac{V' \left( {\phi} \right)^{2} V'''' \left( {\phi} \right)}{V \left( {\phi} \right)^{3}} \right). \\
\end{split}
\end{equation}
\noindent are given by
\begin{equation}
{\epsilon}_{V} = \frac{\left(\chi  F'(\chi )-2 F(\chi )\right)^2}{3 \left(F(\chi )-\chi  F'(\chi) \right)^2}, \\
\end{equation}
\begin{equation}
{\eta}_{V} = -\frac{2 \left(-4 F(\chi ) F''(\chi )+F'(\chi )^2+\chi  F'(\chi ) F''(\chi )\right)}{3 \left(F(\chi )-\chi  F'(\chi )\right) F''(\chi )}, \\
\end{equation}
\begin{equation}
\begin{split}
{\xi}_{V} =& \; \frac{4 \left(\chi  F'(\chi )-2 F(\chi )\right)}{9 \left(F(\chi )-\chi  F'(\chi )\right)^2 F''(\chi )^3} \left[ -8 F(\chi ) F''(\chi)^3+F^{(3)}(\chi ) F'(\chi )^3 \right. \\
& \left. +3 F'(\chi )^2 F''(\chi )^2+\chi  F'(\chi ) F''(\chi)^3\right], \\
\end{split}
\end{equation}
\begin{equation}
\begin{split}
{\omega}_{V} =& \; \frac{-8 \left(\chi  F'(\chi )-2 F(\chi )\right)^2}{27 \left(F(\chi )-\chi  F'(\chi )\right)^3
   F''(\chi )^6} \left[ -18 F(\chi ) F''(\chi )^6 \right. \\
   & \left. +2 F'(\chi ) \left(\chi  F''(\chi )^6+3 F(\chi ) F^{(3)}(\chi ) F''(\chi
   )^4\right) \right. \\
   & \left. +F^{(3)}(\chi ) F'(\chi )^3 \left(2 F(\chi ) F^{(3)}(\chi )^2+2 F''(\chi
   )^3+3 \chi  F^{(3)}(\chi ) F''(\chi )^2\right) \right. \\
   & \left. +F'(\chi )^2 \left(7 F''(\chi )^5-3 \chi  F^{(3)}(\chi ) F''(\chi )^4-6 F(\chi ) F^{(3)}(\chi )^2 F''(\chi
   )^2\right) \right. \\
   & \left. -F'(\chi )^4 \left(\chi  F^{(3)}(\chi )^3+F^{(4)}(\chi ) F''(\chi )^2 - 3 F^{(3)}(\chi )^2 F''(\chi )\right) \right]. \\
\end{split}
\end{equation}

\noindent Then, we can obtain the scalar spectral index ${n}_{s}$ and its running\footnote{In this paper, we adopt ${\simeq}$ to represent slow-roll approximation. }
\begin{equation}
\begin{split}
{n}_{s} \simeq& \; 1 + 2 {\eta}_{V} - 6 {\epsilon}_{V} \\
=& \; \frac{-5 F(\chi )^2 F''(\chi )-2 F(\chi ) F'(\chi ) \left(\chi  F''(\chi )+2 F'(\chi
   )\right)+\chi  F'(\chi )^2 \left(\chi  F''(\chi )+4 F'(\chi )\right)}{3 \left(F(\chi
   )-\chi  F'(\chi )\right)^2 F''(\chi )} \\
\end{split}
\end{equation}
\begin{equation}
\begin{split}
\frac{d {n}_{s}}{d \ln{k}} \simeq& \; 16 {\epsilon}_{V} {\eta}_{V} - 24 {{\epsilon}_{V}}^{2} - 2 {\xi}_{V} \\
=& \; \frac{- 8 \left[ 2 F(\chi )-\chi  F'(\chi) \right]}{9 \left(F(\chi )-\chi  F'(\chi )\right)^4 F''(\chi )^3} \left[ \chi ^2 F'(\chi )^4
   \left(F''(\chi )^2-F^{(3)}(\chi ) F'(\chi )\right) \right. \\
   & \left. +2 \chi  F(\chi ) F'(\chi )^3 \left(F^{(3)}(\chi ) F'(\chi )-3 F''(\chi )^2\right) \right. \\
   & \left. +F(\chi )^2 F'(\chi ) \left(3 \chi  F''(\chi )^3-F^{(3)}(\chi ) F'(\chi )^2+5 F'(\chi ) F''(\chi)^2\right) \right], \\
\end{split}
\end{equation}
\begin{equation}
\begin{split}
\frac{d{n}^{2}_{s}}{d \ln{k}^{2}} \simeq& \; -192 {{\epsilon}_{V}}^{3} + 192 {{\epsilon}_{V}}^{2} {\eta}_{V} - 32 {\epsilon}_{V} {{\eta}_{V}}^{2} - 24 {\epsilon}_{V} {\xi}_{V} + 2 {\eta}_{V} {\xi}_{V} + 2 {\omega}_{V} \\
=& \; \frac{16 \left(\chi  F'(\chi )-2 F(\chi )\right)}{27 \left(F(\chi )-\chi  F'(\chi )\right)^6} \left\{12 \left(2 F(\chi )-\chi  F'(\chi
   )\right)^5 \right. \\
   & \left. -\frac{8 \left(F(\chi )-\chi  F'(\chi )\right)^2 \left(\chi  F'(\chi )-2
   F(\chi )\right) \left(-4 F(\chi ) F''(\chi )+F'(\chi )^2+\chi  F'(\chi ) F''(\chi
   )\right)^2}{F''(\chi )^2} \right. \\
   & \left. -\frac{24 \left(F(\chi )-\chi  F'(\chi )\right) \left(\chi 
   F'(\chi )-2 F(\chi )\right)^3 \left(-4 F(\chi ) F''(\chi )+F'(\chi )^2+\chi  F'(\chi
   ) F''(\chi )\right)}{F''(\chi )} \right. \\
   & \left. -\frac{\left(F(\chi )-\chi  F'(\chi )\right)^3}{F''(\chi )^4}
   \left[ -4 F(\chi ) F''(\chi )+F'(\chi )^2+\chi  F'(\chi ) F''(\chi ) \right] \right. \\
   & \left. \times \left(-8 F(\chi ) F''(\chi )^3+F^{(3)}(\chi ) F'(\chi )^3+3 F'(\chi )^2 F''(\chi )^2+\chi 
   F'(\chi ) F''(\chi )^3\right) \right. \\
   & \left. - \frac{6 \left(F(\chi )-\chi  F'(\chi
   )\right)^2 \left(\chi  F'(\chi )-2 F(\chi )\right)^2}{F''(\chi )^3} \left[-8 F(\chi ) F''(\chi
   )^3+F^{(3)}(\chi ) F'(\chi )^3 \right. \right. \\
   & \left. \left.  +3 F'(\chi )^2 F''(\chi )^2+\chi  F'(\chi ) F''(\chi
   )^3\right] \right. \\
   & \left. +\frac{\left(F(\chi ) - \chi  F'(\chi )\right)^3 \left(2
   F(\chi )-\chi  F'(\chi )\right)}{F''(\chi )^6} \left[ -18 F(\chi ) F''(\chi )^6 \right. \right. \\
   & \left. \left. +2 F'(\chi ) \left(\chi  F''(\chi )^6+3 F(\chi ) F^{(3)}(\chi ) F''(\chi )^4\right) \right. \right. \\
   & \left. \left. +F^{(3)}(\chi ) F'(\chi )^3 \left(2 F(\chi ) F^{(3)}(\chi )^2+2 F''(\chi )^3 + 3 \chi  F^{(3)}(\chi )
   F''(\chi )^2\right) \right. \right. \\
   &\left. \left. +F'(\chi )^2 \left(7 F''(\chi )^5-3 \chi  F^{(3)}(\chi ) F''(\chi
   )^4-6 F(\chi ) F^{(3)}(\chi )^2 F''(\chi )^2\right) \right. \right. \\
   & \left. \left. -F'(\chi )^4 \left(\chi F^{(3)}(\chi )^3+F^{(4)}(\chi ) F''(\chi )^2-3 F^{(3)}(\chi )^2 F''(\chi) \right) \right] \right\}. \\
\end{split}
\end{equation}

\noindent Also, the tensor spectral index and its running are given by
\begin{equation}
\begin{split}
{n}_{t} \simeq& \; - 2 {\epsilon}_{V} = \frac{-2 \left[ \chi  F'(\chi )-2 F(\chi ) \right]^2}{3 \left[ F(\chi )-\chi  F'(\chi) \right]^2}, \\
\frac{d{n}_{t}}{d \ln{k}} \simeq& \; 4 {\epsilon}_{V} {\eta}_{V} - 8 {{\epsilon}_{V}}^{2} \\
=& \; \frac{ -8 \left(\chi  F'(\chi )-2 F(\chi )\right)^2 \left[ F(\chi ) F'(\chi ) \left(\chi 
   F''(\chi )+F'(\chi )\right)-\chi  F'(\chi )^3 \right]}{9 \left(F(\chi )-\chi  F'(\chi
   )\right)^4 F''(\chi )}, \\
\end{split}
\end{equation}

\noindent and the tensor-to-scalar ratio is given by
\begin{equation}
r = \frac{P_{\text{s}}(k)}{P_{\text{t}}(k)} \simeq 16 {\epsilon}_{V} = \frac{16 \left[ \chi  F'(\chi )-2 F(\chi ) \right]^2}{3 \left[ F(\chi )-\chi  F'(\chi) \right]^2}. \\
\end{equation}
\noindent where the scalar and tensor power spectra at a pivot scale $k$, $P_{\text{s}}(k)$ and $P_{\text{t}}(k)$ are given by
\begin{equation}
P_{\text{s}}(k) = \frac{1}{24 \pi^{2} M^{4}_{\text{pl}}} \frac{V}{{\epsilon}_{V}}, \quad P_{\text{t}}(k) = \frac{8}{M^{2}_{\text{pl}}} \left( \frac{H}{2 \pi} \right)^{2}. \\
\end{equation}

\section{Specific derivatives of BI-extended model}
\subsection{The BI-extended model and its derivatives}
\noindent Given that our model is 
\begin{equation}
F(R) = R + \frac{2g^2}{3 {\beta}} \left( \sqrt{1+12 {\beta} R^2} - 1 \right), \\
\end{equation}
\noindent the first, second, third, fourth and fifth derivatives of $F \left( R \right)$ are
\begin{equation}
F'(R) := F^{\left( 1 \right)} \left( R \right) = 1+ \frac{8 g^2 R}{\sqrt{1 + 12 \beta R^2}}, \\
\end{equation}
\begin{equation}
F''(R) := F^{\left( 2 \right)} \left( R \right) = \frac{8 g^2}{\left(12 \beta  R^2 + 1 \right)^{\frac{3}{2}}}, \\
\end{equation}
\begin{equation}
F'''(R) := F^{\left( 3 \right)} \left( R \right) = -\frac{288 \beta  g^2 R}{\left(12 \beta  R^2+1\right)^{5/2}}, \\
\end{equation}
\begin{equation}
F''''(R) := F^{\left( 4 \right)} \left( R \right) = \frac{288 \beta  g^2 \left(48 \beta  R^2-1\right)}{\left(12 \beta  R^2+1\right)^{7/2}}, \\
\end{equation}
\begin{equation}
F'''''(R) := F^{\left( 5 \right)} \left( R \right) = -\frac{51840 \beta ^2 g^2 R \left(16 \beta  R^2-1\right)}{\left(12 \beta R^2+1\right)^{9/2}}. \\
\end{equation}

\subsection{ $\frac{d {\chi}}{d {\phi}}$ and its derivatives}
\noindent $\frac{d {\chi}}{d {\phi}}$ and its derivatives are given by
\begin{equation}
\frac{d {\chi}}{d {\phi}} = \left( \frac{1}{{M}_{\text{pl}}} \sqrt{\frac{2}{3}} \right) \frac{\left(12 \beta  {\chi}^2 + 1\right)}{8 g^2} \left(8 g^2 {\chi} + \sqrt{12 \beta  {\chi}^2 + 1}\right), \\
\end{equation}
\begin{equation}
\frac{d}{d {\chi}} \left( \frac{d {\chi}}{d {\phi}} \right) = \left( \frac{1}{{M}_{\text{pl}}} \sqrt{\frac{2}{3}} \right) \left[ \frac{9 \beta  {\chi}}{2 g^2} \sqrt{12 \beta  {\chi}^2 + 1} + 36 \beta  {\chi}^2 + 1 \right], \\
\end{equation}
\begin{equation}
\frac{d^2}{d {\chi}^2} \left( \frac{d {\chi}}{d {\phi}} \right) = \left( \frac{1}{{M}_{\text{pl}}} \sqrt{\frac{2}{3}} \right) \frac{9 \beta  \left(16 g^2 {\chi} \sqrt{12 \beta  {\chi}^2 + 1} + 24 \beta  {\chi}^2 + 1 \right)}{2 g^2 \sqrt{12 \beta  {\chi}^2 + 1}}, \\
\end{equation}
\begin{equation}
\frac{d^3}{d {\chi}^3} \left( \frac{d {\chi}}{d {\phi}} \right) = \left( \frac{1}{{M}_{\text{pl}}} \sqrt{\frac{2}{3}} \right) 18 \beta  \left[ \frac{9 \beta  {\chi} \left(8 \beta  {\chi}^2 + 1 \right)}{g^2 \left(12 \beta {\chi}^2 + 1 \right)^{3/2}} + 4 \right]. \\
\end{equation}

\subsection{Potential $V$ and its derivatives with respect to $\chi$}
\noindent Based on the specific model, the potential is given by
\begin{equation}
V \left( \chi \right) = \frac{g^2 {M}^{2}_{\text{pl}} \sqrt{12 \beta  \chi ^2+1} \left(\sqrt{12 \beta  \chi ^2+1}-1\right)}{3
   \beta  \left(\sqrt{12 \beta  \chi ^2+1}+8 g^2 \chi \right)^2}, \\
\end{equation}
\begin{equation}
V' \left( \chi \right) = \frac{4 g^2 {M}^{2}_{\text{pl}} \left[ 3 \beta  \chi  \sqrt{12 \beta  \chi ^2+1}-4 g^2 \left(-6 \beta 
   \chi ^2+\sqrt{12 \beta  \chi ^2+1}-1\right)\right]}{3 \beta  \sqrt{12 \beta  \chi
   ^2+1} \left(\sqrt{12 \beta  \chi ^2+1}+8 g^2 \chi \right)^3}, \\
\end{equation}
\begin{equation}
\begin{split}
V'' \left( \chi \right) =& \; \frac{ 4 g^2 {M}^{2}_{\text{pl}} }{\beta  \left(12 \beta  \chi ^2+1\right)^{3/2} \left(\sqrt{12
   \beta  \chi ^2+1}+8 g^2 \chi \right)^4}  \left[ -288 \beta ^3 \chi ^4-12 \beta ^2 \chi ^2+\beta \right. \\
   &\; \left. + 32 g^4 \left( -48 \beta ^2 \chi ^4+6 \beta  \chi ^2 \left(2 \sqrt{12 \beta  \chi^2+1}-3\right)+\sqrt{12 \beta  \chi ^2+1}-1\right) \right. \\
   &\; \left. -16 \beta  g^2 \chi  \left(12 \beta  \chi ^2 \left(2 \sqrt{12 \beta  \chi ^2+1}-3\right)+4 \sqrt{12 \beta  \chi ^2+1}-3\right)\right], \\
\end{split}
\end{equation}
\noindent and $V''' \left( \chi \right)$, $V'''' \left( \chi \right)$ can also be evaluated similarly. 

\subsection{Derivatives of the potential $V$ with respect to $\phi$}
\noindent The derivatives of the potential $V$ with respect to $\phi$ in terms of $\chi$ are given by
\begin{equation}
\frac{d V}{d {\phi}} = \frac{ {M}_{\text{pl}} \sqrt{72 \beta  \chi ^2 + 6} }{18 \beta  \left(\sqrt{12 \beta  \chi^2+1}+8 g^2 \chi \right)^2}
\left\{ 3 \beta  \chi  \sqrt{12 \beta  \chi ^2+1}-4 g^2 \left( -6 \beta  \chi ^2+\sqrt{12 \beta  \chi ^2+1}-1\right) \right\}, \\
\end{equation}
\begin{equation}
\begin{split}
\frac{d^2 V}{d {\phi}^2} =&\; \frac{1}{72 \beta  g^2 \left(\sqrt{12
   \beta  \chi ^2+1}+8 g^2 \chi \right)^2} \left\{ 3 \beta  \left(12 \beta  \chi ^2+1\right)^{5/2} \right. \\
   &\; \left. + 64 g^4 \sqrt{12 \beta  \chi ^2+1} \left(36 \beta ^2 \chi ^4-6 \beta  \chi ^2+\sqrt{12 \beta  \chi ^2+1}-1\right) \right. \\
   &\; \left. +24 \beta  g^2 \chi \left(12 \beta  \chi ^2+1\right) \left(24 \beta  \chi ^2-1\right) \right\}, \\
\end{split}
\end{equation}
\noindent and $\frac{d^3 V}{d {\phi}^3}$, $\frac{d^4 V}{d {\phi}^4}$ can also be evaluated similarly. 

\section{Specific derivatives of Starobinsky model}
\subsection{Derivatives}
\noindent Given that the Starobinsky model is given by
\begin{equation}
F \left( R \right) = R + 4 g^2 R^2, \\
\end{equation}
\noindent the first and second derivatives of $F \left( R \right)$ are
\begin{equation}
F' \left( R \right) := F^{\left( 1 \right)} \left( R \right) = 1 + 8 g^2 R, \\
\end{equation}
\begin{equation}
F'' \left( R \right) := F^{\left( 2 \right)} \left( R \right) = 8 g^2, \\
\end{equation}
\noindent while the derivatives of $F \left( R \right)$ higher than the second order with respect to $R$ vanish. The relationship between $R$ and the inflaton field $\phi$ is given by
\begin{equation}
F' \left( R \right) = {e}^{ \sqrt{\frac{2}{3}} \frac{\phi}{{M}_{\text{pl}}}} \quad \Rightarrow \quad R = \frac{1}{8 g^2} \left( {e}^{\sqrt{\frac{2}{3}} \frac{\phi}{{M}_{\text{pl}}}} - 1\right). \\
\end{equation}

\subsection{ $\frac{d {\chi}}{d {\phi}}$ and its derivatives}
\noindent $\frac{d {\chi}}{d {\phi}}$ and its derivatives are given by
\begin{equation}
\frac{d {\chi}}{d {\phi}} = \left( \frac{1}{{M}_{\text{pl}}} \sqrt{\frac{2}{3}} \right) \frac{1 + 8 g^2 {\chi}}{8 g^2}, \quad \frac{d}{d {\chi}} \left( \frac{d {\chi}}{d {\phi}} \right) = \left( \frac{1}{{M}_{\text{pl}}} \sqrt{\frac{2}{3}} \right), \\
\end{equation}
\noindent while the derivatives of $\frac{d {\chi}}{d {\phi}}$ higher than the first order with respect to $\chi$ vanish. 

\subsection{Potential $V$ and its derivatives with respect to $\chi$}
\noindent Based on the Starobinsky model, the potential and its derivatives are given by
\begin{equation}
V \left( \chi \right) = \frac{2 g^2 \chi ^2 M_{\text{pl}}^2}{\left(8 g^2 \chi +1\right)^2}, \\
\end{equation}
\begin{equation}
V' \left( \chi \right) = \frac{4 g^2 \chi  M_{\text{pl}}^2}{\left(8 g^2 \chi +1\right)^3}, \\
\end{equation}
\begin{equation}
\begin{split}
V'' \left( \chi \right) =& \; -\frac{4 g^2 \left(16 g^2 \chi -1\right) M_{\text{pl}}^2}{\left(8 g^2 \chi +1\right)^4}, \\
\end{split}
\end{equation}
\begin{equation}
\begin{split}
V''' \left( \chi \right) =& \; \frac{192 g^4 \left(8 g^2 \chi -1\right) M_{\text{pl}}^2}{\left(8 g^2 \chi +1\right)^5}, \\
\end{split}
\end{equation}
\begin{equation}
\begin{split}
V'''' \left( \chi \right) =& \; -\frac{3072 g^6 \left(16 g^2 \chi -3\right) M_{\text{pl}}^2}{\left(8 g^2 \chi + 1\right)^6}. \\
\end{split}
\end{equation}

\subsection{Derivatives of the potential $V$ with respect to $\phi$}
\noindent The derivatives of the potential $V$ with respect to $\phi$ in terms of $\chi$ are given by
\begin{equation}
\frac{d V}{d {\phi}} = \frac{d V}{d {\chi}} \frac{d {\chi}}{d {\phi}} = \frac{{M}_{\text{pl}} \chi }{\sqrt{6} \left(8 g^2 \chi +1\right)^2}, \\
\end{equation}
\begin{equation}
\begin{split}
\frac{d^2 V}{d {\phi}^2} =&\; \frac{1-8 g^2 \chi }{24 g^2 \left(8 g^2 \chi + 1 \right)^2}, \\
\end{split}
\end{equation}
\begin{equation}
\begin{split}
\frac{d^3 V}{d {\phi}^3} =&\; \frac{8 g^2 \chi -3}{12 \sqrt{6} g^2 \left(8 g^2 \chi + 1 \right)^2 {M}_{\text{pl}}}, \\
\end{split}
\end{equation}
\begin{equation}
\begin{split}
\frac{d^4 V}{d {\phi}^4} =&\; \frac{7-8 g^2 \chi }{36 g^2 \left(8 g^2 \chi +1\right)^2 M_{\text{pl}}^2}. \\
\end{split}
\end{equation}

\subsection{Slow roll parameters in terms of $\chi$}
\noindent The slow roll parameters in terms of $\chi$ are given by
\begin{equation}
{\epsilon}_{V} \left( {\chi} \right) = \frac{1}{48 g^4 \chi ^2}, \quad {\eta}_{V} \left( {\chi} \right) = \frac{1-8 g^2 \chi }{48 g^4 \chi ^2}, \quad {\xi}_{V}\left( {\chi} \right) = \frac{8 g^2 \chi -3}{288 g^6 \chi ^3}, \quad {\omega}_{V} \left( {\chi} \right) = \frac{7-8 g^2 \chi }{1728 g^8 \chi ^4}. \\
\end{equation}

\noindent The scalar spectral index and its runnings in terms of $\chi$ are given by
\begin{equation}
{n}_{s} \simeq 1 + 2 {\eta}_{V} - 6 {\epsilon}_{V} = -\frac{1}{12 g^4 {\chi}^2}-\frac{1}{3 g^2 {\chi}}+1, \\
\end{equation}
\begin{equation}
\frac{d {n}_{s}}{d \ln{k}} \simeq 16 {\epsilon}_{V} {\eta}_{V} - 24 {\epsilon}^{2}_{V} - 2 {\xi}_{V} = - \frac{16 g^4 \chi ^2+10 g^2 \chi +1}{288 g^8 \chi ^4}, \\
\end{equation}
\begin{equation}
\begin{split}
\frac{d^2 {n}_{s}}{d \ln{k}^2} \simeq& \; - 192 {\epsilon}^{3}_{V} + 192 {\epsilon}^{2}_{V} {\eta}_{V} - 32 {\epsilon}_{V} {\eta}^{2}_{V} - 24 {\epsilon}_{V} {\xi}_{V} + 2 {\eta}_{V} {\xi}_{V} + 2 {\omega}_{V} \\
=& \; -\frac{128 g^6 \chi ^3+136 g^4 \chi ^2+31 g^2 \chi +2}{6912 g^{12} \chi ^6}. \\
\end{split}
\end{equation}
\noindent The scalar power spectrum evaluated at the first horizon crossing is given by
\begin{equation}
\begin{split}
{A}_{s} =& \; {P}_{s} \left( {k}_{\text{hc}} \right) \simeq \frac{1}{ 24 {\pi}^{2} {M}^{4}_{\text{pl}} } \frac{V \left( {\chi}_{\text{hc}} \right)}{{\epsilon}_{V} \left( {\chi}_{\text{hc}} \right)} = \left. \frac{4 g^6 \chi ^4}{\pi ^2 {M}^{2}_{\text{pl}} \left(8 g^2 \chi +1\right)^2} \right|_{\chi = {\chi}_{\text{hc}}}. \\
\end{split}
\end{equation}
\noindent Also, the tensor spectral index and its runnings in terms of $\chi$ are given by
\begin{equation}
{n}_{t} \simeq - 2 {\epsilon}_{V} = -\frac{1}{24 g^4 \chi ^2}, \\
\end{equation}
\begin{equation}
\frac{d {n}_{t}}{d \ln{k}} \simeq 4 {\epsilon}_{V} {\eta}_{V} - 8 {\epsilon}^{2}_{V} = -\frac{8 g^2 \chi +1}{576 g^8 \chi ^4}, \\
\end{equation}
\noindent and the tensor-to-scalar ratio in terms of $\chi$ is given by
\begin{equation}
r \simeq 16 {\epsilon}_{V} = \frac{1}{3 g^4 \chi ^2}. \\
\end{equation}

\subsection{Slow roll parameters in terms of $\phi$}
\noindent The slow roll parameters in terms of $\phi$ are given by
\begin{equation}
\begin{split}
{\epsilon}_{V} \left( {\phi} \right) = \frac{4}{3} \left(e^{\sqrt{\frac{2}{3}} \frac{\phi }{M_{\text{pl}}}}-1\right)^{-2}, \quad&\;
{\eta}_{V} \left( {\phi} \right) = \frac{- 4 \left(e^{\sqrt{\frac{2}{3}} \frac{\phi }{M_{\text{pl}}}}-2\right)}{3 \left(e^{\sqrt{\frac{2}{3}} \frac{\phi }{M_{\text{pl}}}}-1\right)^{2}}, \\
{\xi}_{V} \left( {\phi} \right) = \frac{16 \left(e^{\sqrt{\frac{2}{3}} \frac{\phi }{M_{\text{pl}}}}-4\right)}{9
   \left(e^{\sqrt{\frac{2}{3}} \frac{\phi }{M_{\text{pl}}}}-1\right)^{3}}, \quad&\;
{\omega}_{V} \left( {\phi} \right) = \frac{- 64 \left(e^{\sqrt{\frac{2}{3}} \frac{\phi }{M_{\text{pl}}}}-8\right)}{27
   \left(e^{\sqrt{\frac{2}{3}} \frac{\phi }{M_{\text{pl}}}}-1\right)^{4}}. \\
\end{split}
\end{equation}

\noindent The scalar spectral index and its runnings in terms of $\phi$ are given by
\begin{equation}
{n}_{s} \simeq 1 + 2 {\eta}_{V} - 6 {\epsilon}_{V} = \frac{-14 e^{\sqrt{\frac{2}{3}} \frac{\phi }{M_{\text{pl}}}}+3 e^{2 \sqrt{\frac{2}{3}} \frac{\phi }{M_{\text{pl}}}}-5}{3 \left(e^{\sqrt{\frac{2}{3}} \frac{\phi }{M_{\text{pl}}}}-1\right)^{2}}, \\
\end{equation}
\begin{equation}
\frac{d {n}_{s}}{d \ln{k}} \simeq 16 {\epsilon}_{V} {\eta}_{V} - 24 {\epsilon}^{2}_{V} - 2 {\xi}_{V} = \frac{- 32 e^{\sqrt{\frac{2}{3}} \frac{\phi }{M_{\text{pl}}}}
   \left(e^{\sqrt{\frac{2}{3}} \frac{\phi }{M_{\text{pl}}}}+3\right)}{9 \left(e^{\sqrt{\frac{2}{3}} \frac{\phi }{M_{\text{pl}}}}-1\right)^{4}}, \\
\end{equation}
\begin{equation}
\begin{split}
\frac{d^2 {n}_{s}}{d \ln{k}^2} \simeq& \; - 192 {\epsilon}^{3}_{V} + 192 {\epsilon}^{2}_{V} {\eta}_{V} - 32 {\epsilon}_{V} {\eta}^{2}_{V} - 24 {\epsilon}_{V} {\xi}_{V} + 2 {\eta}_{V} {\xi}_{V} + 2 {\omega}_{V} \\
=& \; \frac{- 128 e^{\sqrt{\frac{2}{3}} \frac{\phi }{M_{\text{pl}}}} \left(11 e^{\sqrt{\frac{2}{3}} \frac{\phi }{M_{\text{pl}}}}+2 e^{2 \sqrt{\frac{2}{3}} \frac{\phi }{M_{\text{pl}}}}+3\right)}{27 \left( e^{\sqrt{\frac{2}{3}} \frac{\phi }{M_{\text{pl}}}} - 1 \right)^{6}}. \\
\end{split}
\end{equation}
\noindent The scalar power spectrum evaluated at the first horizon crossing is given by
\begin{equation}
\begin{split}
{A}_{s} =& \; {P}_{s} \left( {k}_{\text{hc}} \right) \simeq \frac{1}{ 24 {\pi}^{2} {M}^{4}_{\text{pl}} } \frac{V \left( {\phi}_{\text{hc}} \right)}{{\epsilon}_{V} \left( {\phi}_{\text{hc}} \right)} = \left. \frac{1}{1024 \pi ^2 g^2 {M}^{2}_{\text{pl}} } e^{2 \sqrt{\frac{2}{3}} \frac{\phi}{ {M}_{\text{pl}} }} \left(1 - e^{ - \sqrt{\frac{2}{3}} \frac{\phi}{ {M}_{\text{pl}} }} \right)^4 \right|_{ {\phi} = {\phi}_{\text{hc}} }. \\
\end{split}
\end{equation}
\noindent Also, the tensor spectral index and its runnings in terms of $\phi$ are given by
\begin{equation}
{n}_{t} \simeq - 2 {\epsilon}_{V} = \frac{- 8}{3} \left(e^{\sqrt{\frac{2}{3}} \frac{\phi }{M_{\text{pl}}}} - 1 \right)^{- 2}, \\
\end{equation}
\begin{equation}
\frac{d {n}_{t}}{d \ln{k}} \simeq 4 {\epsilon}_{V} {\eta}_{V} - 8 {\epsilon}^{2}_{V} = \frac{- 64}{9} e^{\sqrt{\frac{2}{3}} \frac{\phi }{M_{\text{pl}}}} \left(e^{\sqrt{\frac{2}{3}} \frac{\phi }{M_{\text{pl}}}} - 1\right)^{- 4}, \\
\end{equation}
\noindent and the tensor-to-scalar ratio in terms of $\phi$ is given by
\begin{equation}
r \left( {\phi} \right) \simeq \frac{64}{3} \left(e^{\sqrt{\frac{2}{3}} \frac{\phi}{M_{\text{pl}}}}-1\right)^{-2}. \\
\end{equation}

\section{Data points for the ${n}_{RR} - r$ graph}
\noindent In this part, all the data points $\left( {n}_{s}, r \right)$ (TT,TE,EE+lowE+lensing) with marginalized joint 68\% CL in the ${n}_{s} - r$ graph Fig.(\ref{fig: ns-r graph}) are listed in Table \ref{table: Data points Planck 2018}. 
\begin{table}[h]
\begin{center}
\begin{tabular}{ |c|c|c|c|c|c| }
\hline
$\text{Nothing (Gray)}$ & $\text{BK14 (Red)}$ & $\text{BK14+BAO (Blue)}$ \\
\hline 
$(0.958208,0)$ & $(0.959498,0)$ & $(0.962079,0)$ \\
\hline 
$(0.958638,0.00819672)$ & $(0.95914,0.00819672)$ & $(0.961649,0.00819672)$ \\
\hline 
$(0.95914,0.0163934)$ & $(0.959283,0.0163934)$ & $(0.961505,0.0163934)$ \\
\hline 
$(0.959857,0.026776)$ & $(0.959857,0.026776)$ & $(0.961935,0.026776)$ \\
\hline 
$(0.960645,0.036612)$ & $(0.960789,0.036612)$ & $(0.962724,0.036612)$ \\
\hline 
$(0.961219,0.0445355)$ & $(0.962007,0.0445355)$ & $(0.963799,0.0445355)$ \\
\hline 
$(0.961935,0.05)$ & $(0.963656,0.05)$ & $(0.965233,0.05)$ \\
\hline 
$(0.962939,0.0565574)$ & $(0.966237,0.0543716)$ & $(0.967742,0.0543716)$ \\
\hline 
$(0.964301,0.0642077)$ & $\text{NaN}$ & $\text{NaN}$ \\
\hline 
$(0.966667,0.0688525)$ & $\text{NaN}$ & $\text{NaN}$ \\
\hline 
$(0.96853,0.0642077)$ & $\text{NaN}$ & $\text{NaN}$ \\
\hline 
$(0.969534,0.0565574)$ & $\text{NaN}$ & $\text{NaN}$ \\
\hline 
$(0.970179,0.05)$ & $(0.968244,0.05)$ & $(0.969821,0.05)$ \\
\hline 
$(0.970538,0.0445355)$ & $(0.969391,0.0445355)$ & $(0.970824,0.0445355)$ \\
\hline 
$(0.971039,0.036612)$ & $(0.970251,0.036612)$ & $(0.971541,0.036612)$ \\
\hline 
$(0.97147,0.026776)$ & $(0.970824,0.026776)$ & $(0.972043,0.026776)$ \\
\hline 
$(0.971756,0.0163934)$ & $(0.970824,0.0163934)$ & $(0.972115,0.0163934)$ \\
\hline 
$(0.971971,0.00819672)$ & $(0.970394,0.00819672)$ & $(0.971685,0.00819672)$ \\
\hline 
$(0.972258,0)$ & $(0.969821,0)$ & $(0.970824,0)$ \\
\hline
\end{tabular}
\end{center}
\caption{Data points (TT,TE,EE+lowE+lensing) with marginalized joint 68\% CL in the ${n}_{RR} - r$ graph of Planck 2018}
\label{table: Data points Planck 2018}
\end{table}

\end{document}